\documentclass[aps,prl,twocolumn,superscriptaddress,floatfix,10pt,nofootinbib]{revtex4-1}
\usepackage{graphicx}
\usepackage{latexsym}
\usepackage{amssymb}
\usepackage{amsmath}
\usepackage{amsfonts}
\usepackage{gensymb}
\usepackage{bm}
\usepackage{multirow}
\usepackage{color}
\usepackage{hyperref}

\newcommand{\ii}{\mathrm{i}}
\newcommand{\dd}{\mathrm{d}}

\newcommand{\ket}[1]{|#1 \protect\rangle} 

\newcommand{\refcite}[1]{Ref.\,\cite{#1}}
\newcommand{\eqnref}[1]{Eq.\,(\ref{#1})}
\newcommand{\figref}[1]{Fig.\,\ref{#1}}

\newcommand{\secref}[1]{Sec.\,\ref{#1}}
\newcommand{\appref}[1]{Appendix~\ref{#1}}

\newcommand{\is}{=}

\newcommand{\N}{N}

\newcommand{\A}{\alpha}
\newcommand{\B}{\beta}
\newcommand{\C}{\gamma}
\newcommand{\D}{\delta}

\newcommand{\Hh}{\hat{H}}
\newcommand{\Zh}{\hat{Z}}
\newcommand{\Xh}{\hat{X}}
\newcommand{\Oh}{\hat{\mathcal{O}}}

\newcommand{\Ah}{\hat{\mathcal{A}}}
\newcommand{\Bh}{\hat{\mathcal{B}}}
\newcommand{\Ch}{\hat{\mathcal{C}}}
\newcommand{\Fh}{\hat{\mathcal{F}}}

\newcommand{\Z}{Z}
\newcommand{\X}{X}

\newcommand{\I}{i}
\newcommand{\tI}{\tilde{I}}
\newcommand{\J}{j}

\newcommand{\XX}[1]{\X^{#1}}
\newcommand{\II}[1]{\I_{#1}}

\newcommand{\bx}{\bar{x}}
\newcommand{\by}{\bar{y}}
\newcommand{\bz}{\bar{z}}

\newcommand{\R}{x^\mu}
\newcommand{\tR}{{\tilde{x}^\mu}}
\newcommand{\tRR}{\tR(x^\nu)}

\newcommand{\tT}{{\tilde{t}}}
\newcommand{\tX}{{\tilde{x}}}
\newcommand{\tY}{{\tilde{y}}}
\newcommand{\tZ}{{\tilde{z}}}
\newcommand{\tA}{{\tilde{\A}}}
\newcommand{\tB}{{\tilde{\B}}}
\newcommand{\tC}{{\tilde{\C}}}

\newcommand{\DD}{\D^2}
\newcommand{\DDD}{\D^3}

\newcommand{\del}{\partial}
\newcommand{\eps}{\epsilon}
\newcommand{\dgr}{\dagger}
\newcommand{\nn}{\nonumber}
\newcommand{\vx}{\mathbf{x}}
\newcommand{\Th}{\theta}

\newcommand{\EoM}{\stackrel{\text{EoM}}{=}}

\hypersetup{
  colorlinks = true,
  urlcolor = blue,
  pdfauthor = {Kevin Slagle, Yong Baek Kim},
  pdftitle = {Slagle and Kim - 2017 - Quantum Field Theory of X-Cube Fracton Topological Order and Robust Degeneracy from Geometry}
}

\begin{document}

\title{Quantum Field Theory of X-Cube Fracton Topological Order \texorpdfstring{\\}{} and Robust Degeneracy from Geometry}

\author{Kevin Slagle}
\affiliation{Department of Physics, University of Toronto, Toronto, Ontario M5S 1A7, Canada}

\author{Yong Baek Kim}
\affiliation{Department of Physics, University of Toronto, Toronto, Ontario M5S 1A7, Canada}
\affiliation{Canadian Institute for Advanced Research, Toronto, Ontario, M5G 1Z8, Canada}

\begin{abstract}
We propose a quantum field theory description of the X-cube model of fracton topological order.
The field theory is not (and cannot be) a topological quantum field theory (TQFT),
  since unlike the X-cube model, TQFTs are invariant (i.e. symmetric) under continuous spacetime transformations.
However, the theory is instead invariant under a certain subgroup of the conformal group.
We describe how braiding statistics and ground state degeneracy are reproduced by the field theory,
  and how the the X-cube Hamiltonian and field theory can be minimally coupled to matter fields.
We also show that even on a manifold with trivial topology,
  spatial curvature can induce a ground state degeneracy that is stable to arbitrary local perturbations!
Our formalism may allow for the description of other fracton field theories,
  where the only necessary input is an equation of motion for a charge density.
\end{abstract}


\maketitle

Just as the initial theoretical discovery of (liquid \cite{ZengLiquids}) topologically ordered phases of matter \cite{Wen2D,Lan2017} led to incredible discoveries,
  the same is now occurring in the context of non-liquid topological order, particularly fracton topological order \cite{HaahSelfCorrection,VijayFracton,VijayNonabelian,Brown2016,Bravyi2011,HaahCode,Yoshida2013,MaLayers,PremHaahNandkishore,ChamonModel,WilliamsonUngauging,Slagle2spin,HsiehPartons,Hsieh2017,Prem_Pretko_Nandkishore_2017,Devakul_Parameswaran_Sondhi_2017,Petrova_Regnault_2017}.
Both kinds of topological order have a finite energy gap to all excitations
  (although gapless versions of liquid and non-liquid \cite{PretkoU1,PretkoGravity,Rasmussen2016,Xu2006,Xu2008,PretkoTheta} phases also exist),
  and both host degenerate ground states which are stable to arbitrary perturbations and can only be distinguished by nonlocal operators.
(This is in contrast to spontaneous symmetry breaking states where the degenerate ground states are protected by symmetry and can be distinguished by local order parameters.)
Both liquid and non-liquid topological orders also host topological excitations,
  which can only be annihilated via contact with the appropriate antiparticle(s).

The low energy physics of liquid topologically ordered phases \cite{ZengLiquids} is topologically invariant:
  i.e. symmetric under any continuous (and bijective) spacetime transformation which preserves the topology of the spacetime manifold.
For example, the multiplicity of the ground state degeneracy depends only on the topology of the spatial manifold.
Additionally, the braiding statistics of the topological excitations depend only on the topology of the paths that they take.
Exactly solvable lattice models exist for many of these phases \cite{LevinWenModel}.
However, the topological nature of these phases lends to a more minimal and universal description in the form of a topological quantum field theory (TQFT)
  \cite{ChernSimons,BartlettCategorical,SchwarzTQFT,WittenTQFT,AtiyahTQFT,dijkgraafWitten}
  which makes the topological nature of these phases explicit via an explicit topological invariance,
  which is not possible in a lattice model.
For example, Kitaev's toric code lattice model and the BF theory TQFT in 2+1D (or equivalently Chern-Simons theory with $K$-matrix \cite{WenKMatrix} $K=2\sigma^x$) both describe $Z_2$ topological order \cite{KitaevToric,BFTheory,PutrovBraiding}.

Non-liquid topologically ordered phases retain many of the exotic properties of the liquid topological phases,
  except that the long distance physics is not topologically invariant.
The simplest example of a non-liquid phase is a decoupled (or weakly coupled) stack of 2D toric codes in three dimensions.
A less trivial example is the X-cube model \cite{VijayXCube} with fracton topological order which we study in this work.
Both models have a ground state degeneracy which is stable to arbitrary perturbations;
  however, on an $L \times L \times L$ torus their degeneracy is exponentially large with $L$,
  and thus depends on more than just the topology of the spatial manifold.
Both phases also have constrained dynamics which are stable to perturbations.
In the stack of toric codes, the topological charge and flux excitations can not move between different toric code stacks,
  which allows the point-like charge and flux particles to have non-trivial braiding statistics in a 3D phase.
And the X-cube model has dimension-1 particles \cite{VijayFracton} which can only move in the $x$, $y$, or $z$ directions.
These movement constraints are not invariant under spatial deformations,
  and thus these phases aren't topologically invariant and therefore can't be described by a TQFT.

Nevertheless, it seems important to ask if it is possible to describe these non-liquid topological phases with a field theory which captures as much of the spacetime symmetry as possible
  (i.e. some subgroup of the group of continuous spacetime transformations).
However, as is often the case in quantum field theories, an obstruction presents itself in the form of an infinite quantity.
In the case of the stacked toric code example, a natural field theory is BF theory with an extra $z$ coordinate to index the different stacks:
\begin{align}
  L_\text{BF stack} &= \frac{1}{\pi} \sum_{\A\B\C=0,1,2} \epsilon^{\A\B\C} B_\A(t,x,y,z) \del_\B A_\C(t,x,y,z) \nn
\end{align}
However, this field theory appears to have an infinite degeneracy on a torus:
  a factor of 4 for each value of $z$, for which there are infinitely many.
Nevertheless, we argue that this dilemma can be solved by applying the standard weapon against infinities in field theories:
  a short distance cutoff.
That is, if we impose a short distance cutoff $a$,
  then the degeneracy on an $l \times l \times l$ torus is finite and equal to $\sim 4^{l/a}$.
Thus, we can view the field theory as describing a periodic $L \times L \times L$ cubic lattice with degeneracy $4^L$ where $L \sim l/a$.
We propose that the same trick can be applied to our field theory for the X-cube model,
  which also has a degeneracy which is exponentially large in system size.

A more practical concern is: how do we write down the correct field theory to describe a given non-liquid phase?
This task was manageable for TQFTs because the topological invariance greatly limited the possible Lagrangians that could be written down.
For the case of non-topologically invariant (i.e. non-liquid) topological phases, we do not have this luxury.

However, for the case of toric code and its BF theory TQFT description,
  the terms in the toric code Hamiltonian can be precisely related to the terms in the BF Lagrangian and also to the gauge invariance of the TQFT.
That is, the time components of the gauge fields act as Lagrange multipliers which impose zero charge and flux constraints,
  where the terms in the toric code Hamiltonian are lattice discretizations of the charge and flux densities in BF theory.
And the gauge invariance is related to the fact that all of the terms in the toric code Hamiltonian commute with each other.
We review this relationship in \appref{app:BF} (and \appref{app:BF 3+1D} for 3+1D BF theory).

In \secref{sec:Xcube} we use this intuition to systematically derive a field theory (\eqnref{eq:L Xcube}) for the $Z_\N$ X-cube model \cite{VijayXCube} of fracton topological order.
The precise relations discussed in the previous paragraph continue to hold.
The field theory is not topologically invariant,
  but is instead invariant under a certain subgroup of the conformal group of spacetime transformations which transforms all coordinates independently (\secref{sec:Xcube invariance}).
However, there are new surprises which challenged our previous intuition.
For example, we will see that the parallel movement of a pair of fractons requires a fracton dipole current in the region between the pair of fractons (\figref{fig:fractonCurrent}),
  which is related to the fact that in the lattice model a membrane operator can be used to move a pair of fractons.
This peculiarity is necessitated by the exotic charge conservation constraints (\eqnref{eq:Xcube constraint})
  which enforce e.g. the immobility of isolated fractons. 

In \secref{sec:Xcube braiding} we explain the generic braiding processes of the X-cube model and how they are described by our field theory.
In particular, the motion of dimension-1 particles around the edges of a cube results in a phase factor which depends on the number of fractons within the cube
  (modulo $\N$ for the $Z_\N$ model) \cite{foot:MaLayer}.
And the motion of a pair of oppositely charged fractons around the top and bottom edge of a cylinder oriented along the $z$-axis generates a phase
  which depends on the difference in the number of $x$-axis and $y$-axis dimension-1 particles \cite{MaLayers}.

In \secref{sec:Xcube matter} we show how the X-cube Hamiltonian and field theory can be coupled to matter fields with subdimensional symmetries.
Before the matter fields are coupled to the gauge fields,
  the matter excitations have the same mobility constraints as the fractons and dimension-1 particles of the X-cube model.
However, while the mobility constraints of the X-cube model are robust (i.e. stable under arbitrary local perturbations),
  the mobility constraints of the matter fields in the absence of the gauge fields is instead protected by subdimensional symmetry.

In \secref{sec:Xcube degen} we explain how the ground state degeneracy of the X-cube model can be calculated from either the lattice model or the field theory.
As is well known, the degeneracy is not topologically invariant, but is instead exponentially large with system size (on a torus) \cite{VijayXCube,VijayLayer,MaLayers}.
However, when the log of the degeneracy is expressed as an integral over space (\eqnref{eq:Xcube degen integral}),
  it is invariant under the spacetime transformation discussed in \secref{sec:Xcube invariance} when the cutoff is transformed appropriately.
We then ask: since the degeneracy isn't topologically invariant, is a nontrivial topology of the spatial manifold actually necessary for a stable ground state degeneracy in the X-cube model?
In \secref{sec:curvature degeneracy} we show that the answer is no:
  a cubic lattice with curvature defects can host a stable ground state degeneracy.
As the size of the curved portion of the lattice increases,
  the degeneracy can be made exponentially large
  while the energy splitting of the degeneracy due to perturbations becomes exponentially small.

In \appref{app:new models} we attempt to use our field theory generating formalism to derive field theories for new non-liquid topological phases.
With our current formalism, the only necessary input is an equation for a charge density.
Unfortunately, in this work we only rule out certain simple possibilities.
For example, we find that when the $U(1)$ scalar charge fracton phase \cite{Rasmussen2016,PretkoU1} is ``Higgsed'' down to $Z_\N$,
  that the fractons in the $U(1)$ theory become mobile (and thus not fractons) in the $Z_\N$ theory.
Other possibilities are left to future work.

\section*{Notation}

Before we begin, we will briefly explain some of the nonstandard notation that we use.
Roman letters $a,b,c,d=1,2,3$ denote spatial indices.
Greek letters $\A,\B,\C,\D=0,1,2,3$ denote spacetime indices.
($a,b,c,d=1,2$ and $\A,\B,\C,\D=0,1,2$ for the 2+1D theories in \appref{app:BF}).
$0$ is the time index.
Hats are placed above operators.
A semicolon (e.g. in $A_{0;a}$ in \secref{sec:Xcube}) is used to indicate that the indices following the semicolon do not transform under spacetime transformations (\secref{sec:Xcube invariance}).

We use the convention that all spatial and spacetime indices are implicitly summed unless they appear on both sides of the equation or the right hand side is zero.
For example,
\begin{align}
  \J^{0;a} &\EoM \frac{\N}{2\pi} \eps^{0abc} \del_c \XX{c} \tag{\ref{eq:Xcube EoM}}\nn\\
  J^b &= 0 \nn \tag{\ref{eq:z-anti}}\nn
\end{align}
could be written more explicitly as
\begin{align}
  \forall_a: \J^{0;a} &\EoM \frac{\N}{2\pi} \sum_{b,c=1,2,3} \eps^{0abc} \del_c \XX{c} \nn\\
  \forall_b: J^b      &= 0 \nn\\
                      &\text{or} \nn\\
             J^{1}    &= J^{2} = J^{3} = 0 \nn
\end{align}
where $\forall_a$ means for all $a=1,2,3$.
The semicolon does not denote a covariant derivative.
Instead, in $\J^{0;a}$ the $0$ is used to emphasize that $\J^{0;a}$ transforms like a time-component of the current $\J$,
  while ``$a$'' mereley indexes the different time-components of $\J^{0;a}$;
  the semicolon is used to seperate these different kinds of indices.

\section{X-Cube Quantum Field Theory}
\label{sec:Xcube}

\subsection{Derivation}

We will begin by systematically deriving a quantum field theory (QFT) for the X-cube model of fracton topological order \cite{VijayXCube}.
See \appref{app:BF derivation} and \appref{app:BF 3+1D}, for analogous derivations for BF theory in 2+1D and 3+1D.

\begin{figure}
\includegraphics[width=.75\columnwidth]{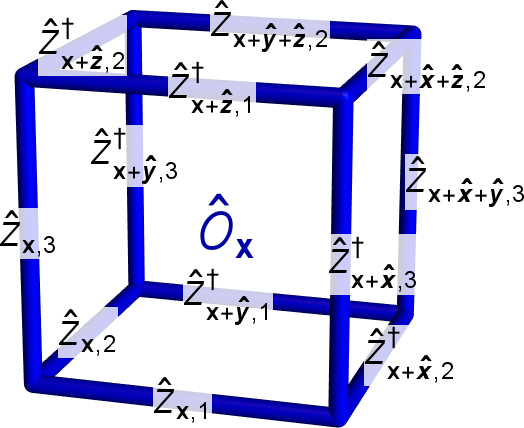}
\includegraphics[width=.32\columnwidth]{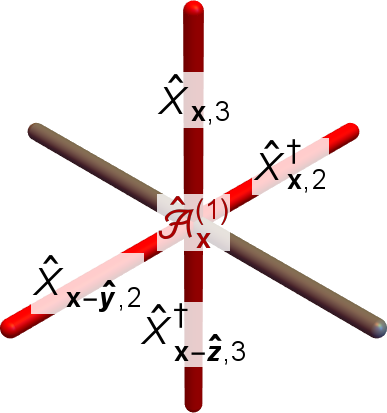}
\includegraphics[width=.32\columnwidth]{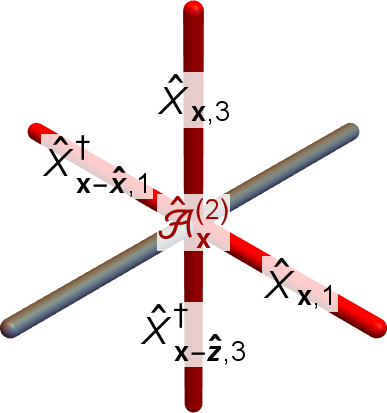}
\includegraphics[width=.32\columnwidth]{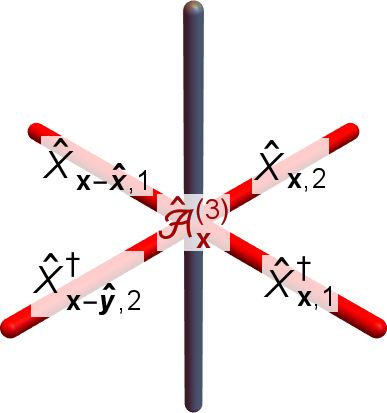}
\caption{
Fracton operator $\Oh$ and dimension-1 particle $\Ah^{(a)}$ operators of the X-cube model (\eqnref{eq:Xcube H}).
$\Oh$ is a product of twelve $\Zh$ operators on the links bordering a cube.
$\Ah^{(a)}$ operators are a product of four $\Xh$ operators on the four links neighboring a vertex which are orthogonal to the $x^a$ direction.
$\Xh$ and $\Zh$ are defined in \eqnref{eq:commutator}.
}\label{fig:Xcube}
\end{figure}

The $Z_\N$ X-cube model is defined by the following Hamiltonian \cite{VijayXCube}:
\begin{align}
  \Hh_\text{X-cube} &= - \sum_\vx (\Oh_\vx + \Oh_\vx^\dag)
                       - \sum_{\vx,a} (\Ah^{(a)}_\vx + \Ah^{(a)\dag}_\vx) \label{eq:Xcube H}
\end{align}
where $\vx=(x,y,z)$ denotes the spatial coordinates.
$\Oh$ and $\Ah^{(a)}$ are defined in \figref{fig:Xcube}
  in terms of $\Zh$ and $\Xh$, which are $Z_\N$ generalizations of Pauli operators:
\begin{align}
  \Xh_i \Zh_j &= \omega^{\D_{ij}} \Zh_j \Xh_i \label{eq:commutator}\\
  \omega &= e^{2\pi \ii/\N} \nn\\
  \text{eigenvalues}&(\Zh_i) = \text{eigenvalues}(\Xh_i) = 1, \omega, \omega^2, \cdots, \omega^{\N-1} \nn
\end{align}
If $\N=2$, then $\Zh$ and $\Xh$ reduce to the usual Pauli operators $\hat\sigma^z$ and $\hat\sigma^x$.
$\Oh = e^{2\pi n/\N}$ is the fracton operator, where $n$ is the number of fracton excitations module $N$.
$\Ah^{(a)}$ is a dimension-1 particle operator;
  if $\N=2$ and $-\Ah^{(1)}_\vx = -\Ah^{(2)}_\vx = \Ah^{(3)}_\vx = 1$,
  then there is a z-axis dimension-1 particle at $\vx$. (\figref{fig:toricCode})

In order to connect the lattice model to the field theory,
  we will rewrite the lattice operators as exponents of fields $\Z_a$ and $\XX{a}$: 
\begin{align}
  \Zh_{\vx,a}(t)   &\sim \exp\biggl( \ii \int'_a                    \Z_a(t,\vx) \biggr) \label{eq:Xcube logZ}\\
  \Xh_{\vx,a}(t)   &\sim \exp\biggl( \ii \int''_{\perp a}         \XX{a}(t,\vx) \biggr) \nn\\
  \Oh_\vx(t)       &\sim \exp\biggl( \frac{2\pi \ii}{\N} \int'      \I^0(t,\vx) \biggr) \nn\\
  \Ah^{(a)}_\vx(t) &\sim \exp\biggl( \frac{2\pi \ii}{\N} \int'' \J^{0;a}(t,\vx) \biggr) \nn
\end{align}
where $a=1,2,3$ is a spatial index (roman letters $a,b,c,d\dots$ are used to denote spatial indices).
$\I^0$ and $\J^{0;a}$ are fracton and dimension-1 particle densities,
  respectively.
For the purposes of this work, we will only interpret \eqnref{eq:Xcube logZ} as a rough correspondence.
The integrals integrate over small regions near $\vx$.
Specifically:
  $\int'_a$ is an integral across the link that $\Zh_{\vx,a}$ lives on;
  $\int''_{\perp a}$ integrates over the dual plaquette that is orthogonal to the link that $\Xh_{\vx,a}$ lives on;
  $\int'$ integrates over the cube that $\Oh_\vx$ is centered at; and
  $\int''$ integrates over the a plaquette in the place of the $\Ah_\vx$ operator.

We will usually view $\Z$ and $\X$ as real-valued fields,
  which are distinguished from their corresponding operators $\Zh$ and $\Xh$ by hats.
However, when $\Z$ and $\X$ are viewed as operators, they have the following equal time commutation relation:
\begin{align}
  [\Z_a(t,\vx), \XX{b}(t,\vx')] &= \frac{2\pi \ii}{\N} \D^b_a \DDD(\vx-\vx') \label{eq:3d bracket'}
\end{align}

Using \eqnref{eq:Xcube logZ}, the fracton and dimension-1 particle densities $\I^0$ and $\J^{0;a}$ can be read off from \figref{fig:Xcube}:
\begin{align}
  \I^0     &\EoM \frac{\N}{2\pi} |\eps^{0abc}| \frac{1}{2} \del_a \del_b \Z_c \label{eq:Xcube EoM}\\
  \J^{0;a} &\EoM \frac{\N}{2\pi} \eps^{0abc} \del_c \XX{c} \nn
\end{align}
where ``$\EoM$'' is used to emphasized that these will be equations of motion
  and not strict equalities.
$\Oh$ and $\Ah^{(a)}$ (\figref{fig:Xcube}) can be viewed as lattice discretizations of $\I^0$ and $\J^{0;a}$.
Note that $\I^0$ and $\J^{0;a}$ commute (i.e. $[\I^0(t,\vx),\J^{0;a}(t,\vx')] = 0$ via the bracket in \eqnref{eq:3d bracket'});
  this occurs because $\Oh$ and $\Ah^{(a)}$ commute (i.e. $[\Oh_\vx,\Ah^{(a)}_{\vx'}] = \Oh_\vx \Ah^{(a)}_{\vx'} - \Ah^{(a)}_{\vx'} \Oh_{\vx} = 0$).

Regarding the notation, $|\eps^{0bcd}|$ is just the absolute value of the Levi-Civita symbol,
  and merely forces $b$, $c$, and $d$ to be different spacetime indices.
We will use the convention that all spatial and spacetime indices are implicitly summed unless they appear on both sides of the equation or the right hand side is zero.
Thus, in the equation for $\J^{0;a}$, $a$ is not summed,
  but both $b$ and $c$ are implicitly summed over even though $b$ only appears once and $c$ appears three times.
The semicolon is used to indicate that the indices following the semicolon do not transform under spacetime transformations (\secref{sec:Xcube invariance}).

The Lagrangian description of the degenerate ground state manifold can now be written down:
\begin{align}
  \tilde{L}_\text{X-cube} &= \frac{\N}{2\pi} \XX{a} \del_0 \Z_a \label{eq:L Xcube'}\\
    &+ \X_0 \underbrace{\frac{\N}{2\pi} |\eps^{0abc}| \frac{1}{2} \del_a \del_b \Z_c}_{\I^0}
     + \Z_{0;a} \underbrace{\frac{\N}{2\pi} \eps^{0abc} \del_c \XX{c}}_{\J^{0;a}} \nn\\
    &- \Z_{0;a} \J^{0;a} - \Z_a \J^a - \X_0 \I^0 - \XX{a} \II{a} \nn\\
    \sum_a \Z_{0;a} &= 0 \label{eq:L Xcube Hspace'}
\end{align}
The first term describes the equal-time commutation relation (\eqnref{eq:3d bracket'}),
  while the second and third terms enforce a zero charge constraint (\eqnref{eq:Xcube EoM})
  via the Lagrange multipliers $\Z_{0;a}$ and $\X_0$.
The final four terms are generic couplings of the fields ($\Z$ and $\X$) to the current sources ($\J$ and $\I$).
\eqnref{eq:L Xcube Hspace'} is a local Hilbert space constraint,
  which results from the fact that $\sum_a \J^{0;a} \EoM 0$ (\eqnref{eq:Xcube EoM}) and $\prod \Ah^{(a)} = 1$.

In order to transition to a more standard notation,
  we will redefine $\Z$, $\X$, $\J$, and $\I$ in term of $A$, $B$, $J$, and $I$, respectively
\footnote{In case the reader is interested, a previous version of this work \cite{firstQFT} was written up using $Z$ and $X$ fields instead,
  which are more closely connected to the Pauli operators $\Zh$ and $\Xh$.}:
\begin{align}
  A_{0;a} &= \Z_{0;a} \label{eq:redef}
& J^{0;a} &= \J^{0;a} \nn\\
  A_a     &= \Z_a
& J^a     &= \J^a \nn\\
  B_0     &= \X_0
& I^0     &= \I^0 \\
  B_{ab}  &= |\epsilon_{0abc}|\, \XX{c}
& I^{ab}  &= |\epsilon^{0abc}|\, \II{c} \nn\\
  \XX{a}  &= |\epsilon^{0abc}|\, \frac{1}{2} B_{bc}
& \II{a}  &= |\epsilon_{0abc}|\, \frac{1}{2} I^{bc} \nn
\end{align}
In terms of the $A$ and $B$ fields, the Lagrangian $\tilde{L}_\text{X-cube}$ (\eqnref{eq:L Xcube'}) becomes:
\begin{align}
  L_\text{X-cube} &= \frac{\N}{2\pi} |\epsilon^{0abc}| \frac{1}{2} B_{ab} \del_0 A_c \label{eq:L Xcube}\\
    &+ B_0 \underbrace{\frac{\N}{2\pi} |\eps^{0abc}| \frac{1}{2} \del_a \del_b A_c}_{I^0}
     + A_{0;a} \underbrace{\frac{\N}{2\pi} \eps^{0abc} \del_c B_{ab}}_{J^{0;a}} \nn\\
    &- A_{0;a} J^{0;a} - A_a J^a - B_0 I^0 - B_{ab} \frac{1}{2} I^{ab} \nn\\
  \sum_a A_{0;a} &= 0 \label{eq:L Xcube Hspace}
\end{align}

The equations of motion for the currents are
\begin{align}
  I^0     &\EoM   \frac{\N}{2\pi} |\eps^{0abc}| \frac{1}{2} \del_a \del_b A_c \label{eq:Xcube EoM'}\\
  I^{ab}  &\EoM   \frac{\N}{2\pi} |\eps^{0abc}| \del_0 A_c - \frac{\N}{2\pi} \eps^{0abc} \del_c (A_{0;a} - A_{0;b}) \nn\\
  J^{0;a} &\EoM   \frac{\N}{2\pi}  \eps^{0abc}  \del_c B_{ab} \nn\\
  J^a     &\EoM - \frac{\N}{2\pi} |\eps^{0abc}| \frac{1}{2} (\del_0 B_{bc} - \del_b \del_c B_0) \nn
\end{align}

The gauge invariance can be derived as follows:
\begin{align}
  B_{ab}(t,\vx) &\rightarrow B_{ab}(t,\vx) \label{eq:Xcube gauge}\\
    &+\ii \int_{\vx'} [B_{ab}(t,\vx),
       \underbrace{\frac{\N}{2\pi} |\eps^{0cde}| \frac{1}{2} \del'_c \del'_d A_e(t,\vx')}_{I^0(t,\vx')}] \chi(t,\vx') \nn\\
    &= B_{ab} + \del_a \del_b \chi \nn\\
  A_a(t,\vx) &\rightarrow A_a(t,\vx) \nn\\
    &+\ii \int_{\vx'} [A_a(t,\vx), \underbrace{\frac{\N}{2\pi} \eps^{0bcd} \del'_d B_{bc}(t,\vx')}_{J^{0;b}(t,\vx')}] \zeta_b(t,\vx') \nn\\
    &= A_a - \eps^{0abc} \del_a \zeta_c \nn\\
  \sum_a \zeta_a &= 0 \label{eq:zeta constraint}
\end{align}
where the brackets $[\cdots,\cdots]$ are evaluated using \eqnref{eq:3d bracket'} written in terms of $A$ and $B$ fields:
\begin{align}
  [A_a(t,\vx), B_{bc}(t,\vx')] &= \frac{2\pi \ii}{\N} |\eps_{0abc}| \DDD(\vx-\vx') \label{eq:3d bracket}
\end{align}
A constraint on $\zeta_a$ (\eqnref{eq:zeta constraint}) is imposed since it does not reduce the generality of the gauge transformation,
  and because it will be needed to fulfill the constraint on $A_{0;a}$ (\eqnref{eq:L Xcube Hspace}) under its gauge transformation (\eqnref{eq:Xcube gauge'}).
The transformation of the fields ($A_a$ and $B_{bc}$) corresponds to conjugating the lattice operators ($\Zh_{\vx,a}$ and $\Xh_{\vx,d}$) by the terms in the Hamiltonian ($\Ah^{(e)}_{\vx'}$ and $\Oh_{\vx'}$)
  at the positions where $\zeta_e(t,\vx')$ and $\chi(t,\vx')$ are nonzero:
  e.g. $\Zh_{\vx,a} \rightarrow \Ah^{(e)\dag}_{\vx'} \Zh_{\vx,a} \Ah^{(e)}_{\vx'}$.
The gauge invariance is a direct result of the fact that the terms in $\Hh_\text{X-cube}$ (\eqnref{eq:Xcube H}) commute with each other.
For example, $I^0$ and $J^{0;a}$ are invariant under the above transformation because
  $I^0$ and $J^{0;a}$ commute (i.e. $[I^0(t,\vx),J^{0;a}(t,\vx')] = 0$),
  and $I^0$ and $J^{0;a}$ commute because $\Oh$ and $\Ah^{(a)}$ commute.

To derive how $A_{0;a}$ and $B_0$ transform,
  the above gauge transformations can be inserted into $L_\text{X-cube}$ (\eqnref{eq:L Xcube}),
  and $A_{0;a}$ and $B_0$ can be solved for to find:
\begin{align}
  A_{0;a} &\rightarrow A_{0;a} + \del_0 \zeta_a \label{eq:Xcube gauge'}\\
  B_0     &\rightarrow B_0     + \del_0 \chi \nn
\end{align}

Finally, in order for the coupling of the gauge fields ($A$ and $B$) to the currents ($I$ and $J$) in $L_\text{X-cube}$ (\eqnref{eq:L Xcube}) to be gauge invariant,
  the currents must obey the following constraint:
\begin{align}
  \del_0 I^0 - \frac{1}{2} \del_a \del_b I^{ab} = 0 \label{eq:Xcube constraint}\\
  \forall_a: \del_0 J^{0;a} + \eps^{0abc} \del_c J^c = 0 \nn
\end{align}
where the ``$\forall_a$'' means that we don't sum over $a$ in the last equation,
  which specifies three separate constraints.
These are generalized charge conservation constraints (analogous to \eqnref{eq:BF constraint} for BF theory),
  which encode the movement restrictions of the fracton current $I$ and dimension-1 particle current $J$.
\footnote{\eqnref{eq:Xcube constraint} is similar to the generalized continuity equation in \refcite{electromagnetismPretko}.}

\subsubsection{Example Currents}

\begin{figure}
\hspace*{1cm}
\includegraphics[width=.6\columnwidth]{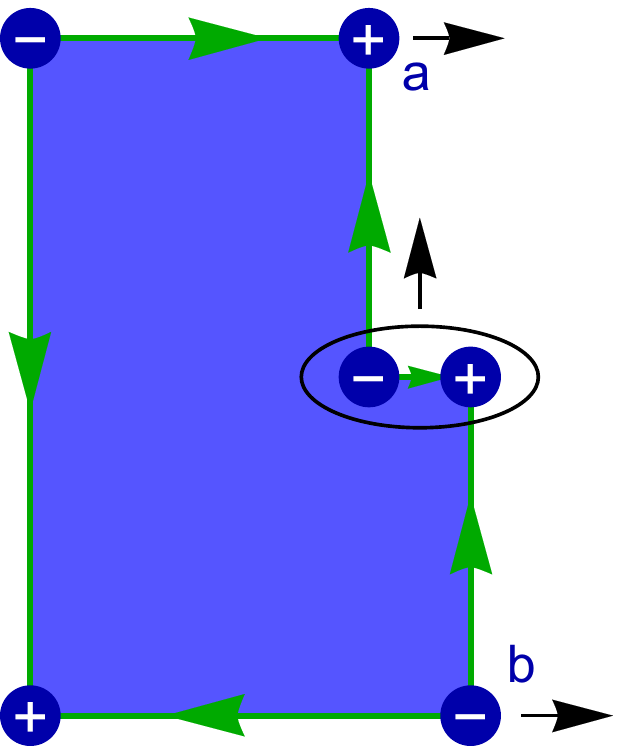}
\caption{
In order for the fractons (labeled a and b) to move to the right,
  they must exchange a fracton dipole (circled in black).
Generating this fracton configuration from the vacuum requires a nonzero fracton current $I^{12}$ in the blue region.
Thus, $I^{cd}$ can be interpreted as a fracton dipole current since it describes a combination of
  1) fracton dipoles oriented in the $x^c$-direction moving in the $x^d$-direction and
  2) fracton dipoles oriented in the $x^d$-direction moving in the $x^c$-direction.
(Recall that $I^{cd} = I^{dc}$.)
The dipole exchange results in a fracton flow (green, \eqnref{eq:fracton flow}),
  most of which will be canceled out by an additional dipole exchange.
}\label{fig:fractonCurrent}
\end{figure}

A single stationary fracton at the origin is simply described by
\begin{align}
  I^0    &= \D(x)\D(y)\D(z) \label{eq:dim1 I}\\
  I^{ab} &= 0 \nn
\end{align}
where $\D(x)$ is the Dirac delta function.
However, the current source ($I$) describing the creation of fractons is more exotic.
\eqnref{eq:Xcube constraint} allows four fractons to be created at $t=0$ and $\vx = (\pm\ell,\pm\ell,0)$, forming a fracton quadrupole, via the following current configuration
\begin{align}
  I^0    &= \sum_{\mu,\nu=\pm1} \mu \nu\, \Th(t)\D(x-\mu \ell)\D(y-\nu \ell)\D(z) \\
  I^{12} &= \D(t) \Th(\ell + x) \Th(\ell - x) \Th(\ell + y) \Th(\ell - y) \D(z) \nn\\
  I^{23} = I^{13} &= 0 \nn
\end{align}
where $\Th(x)$ is the Heaviside step function:
\begin{align}
  \Th(x) = \begin{cases} 0 & x  <  0 \\
                         1 & x \ge 0
           \end{cases} \nn\\
  \del_x \Th(x) = \D(x) \nn
\end{align}
$I^{12}$ is nonzero on a square at time $t=0$,
  which then creates four fractons ($I^0$) at the corners for $t>0$.
This is analogous to the X-cube lattice model where fractons are created at the corners of membrane operators.
The double derivative in $\del_a \del_b I^{ab}$ in \eqnref{eq:Xcube constraint} is the reason why fractons are created at corners of membrane operators in the field theory
  (instead of at the ends of string operators as is typically the case).
Physically, $I^{ab}$ is can be regarded as a fracton dipole current;
  see \figref{fig:fractonCurrent}.

A $z$-axis dimension-1 particle at the origin is represented by
\begin{align}
  J^{0;1} = -J^{0;2} &= \D(x) \D(y) \D(z) \\
  J^{0;3} = J^a &= 0 \nn
\end{align}
This is similar to the lattice model where a $z$-axis particle excites the $\Ah^{(1)}$ and $\Ah^{(2)}$ operators.
The $z$-axis particle can only move in the $z$ direction.
This motion is described by
\begin{align}
  J^{0;1} = -J^{0;2}  &= \phantom{v}\, \D(x) \D(y) \D(z - v t) \\
  J^3                 &= v          \, \D(x) \D(y) \D(z - v t) \nn\\
  J^{0;3} = J^1 = J^2 &= 0 \nn
\end{align}
More generally, an $x^a$-axis particle at the origin is given by
\begin{align}
  & x^a\text{-axis dim-1 particle:} \nn\\
  J^{0;b} &= \sum_c \eps^{0abc} \D(x) \D(y) \D(z) \label{eq:a-axis particle}\\ 
  J^b &= 0 \nn
\end{align}

If both an $x$-axis and $y$-axis dimension-1 particle are at the origin,
  then this is equivalent to a $z$-axis antiparticle:
\begin{align}
  J^{0;b} &= \sum_{a=1,2} \sum_{c=1,2,3} \eps^{0abc} \D(x) \D(y) \D(z) \label{eq:z-anti}\\
          &= - \eps^{03bc} \D(x) \D(y) \D(z) \nn\\
  J^b &= 0 \nn
\end{align}
The first line in \eqnref{eq:z-anti} is \eqnref{eq:a-axis particle} summed over $a=1,2$;
  this corresponds to the presence of both an $x$-axis and a $y$-axis particle.
The second line shows that this is equivalent to the negation of \eqnref{eq:a-axis particle} with $a=3$,
  which corresponds to just a single $z$-axis antiparticle.
(This fusion rule can also be understood from the lattice operators.)

\subsection{``Braiding'' Statistics}
\label{sec:Xcube braiding}

If no additional excitations are created,
  isolated fractons are immobile and isolated dimension-1 particles can only move along straight lines.
However, when we consider braiding statistics of topological excitations,
  we are allowed to create additional excitations.
For example, to measure a flux in $Z_2$ toric code we imagine 1) creating two $Z_2$ charges,
  2) moving one of the charges around the flux,
  and then 3) annihilating the two charges.
We will find a slightly more exotic scenario for the ``braiding'' of X-cube excitations.

\subsubsection{Dimension-1 Particle ``Braiding''}

As a first example, we will use our field theory description to demonstrate that in order to count the number of fractons (modulo $\N$) within a cube,
  we can create dimension-1 particles and move them around the edges of the cube \cite{foot:MaLayer}.

The fracton current $I$ that describes the presence of a single fracton at $\vx=0$ is
\begin{align}
  I^0    &= \DDD(\vx) \tag{\ref{eq:dim1 I}}\\
  I^{ab} &= 0 \nn
\end{align}
A solution to \eqref{eq:Xcube EoM'} to describe this (motionless) current is
\begin{align}
  A_3 = \frac{2\pi}{\N} \Th(x) \Th(y) \D(z) \label{eq:dim1 Z sol}\\
  A_{0;a} = A_1 = A_2 = 0 \nn
\end{align}
Using \eqnref{eq:Xcube logZ}, this can be interpreted as $\Zh_{\vx,3} \sim e^{2\pi \ii/\N}$ (at the mean field level
  \footnote{
    For example, a mean field wavefunction can be defined by $\Zh_{\vx,a} \ket{\psi_\text{MF}} = \exp\left( \ii \int'_a A_a(t,\vx) \right) \ket{\psi_\text{MF}}$
      (e.g. with $A$ given in \eqnref{eq:dim1 Z sol}).
    The physical wavefunction $\ket{\psi}$ is the mean field wavefunction projected onto the desired dimension-1 particle charge configuration:
      $\ket{\psi} = \hat{\mathcal{P}} \ket{\psi_\text{MF}}$
      where the projection operator
      $\hat{\mathcal{P}} = \prod_{\vx,a} \frac{1}{2} \left( \Ah^{(a)}_\vx + \bar{\mathcal{A}}^{(a)}_\vx \right)$ (if $\N=2$)
      projects onto the charge configuration $\bar{\mathcal{A}}^{(a)}_\vx = \pm1$. 
    In \eqnref{eq:dim1 J}, the charge configuration is zero ($J^{0;a} = 0$),
      which corresponds to $\bar{\mathcal{A}}^{(a)}_\vx = 1$.
    If we consider $A_a = 0$ (instead of \eqnref{eq:dim1 Z sol}) and keep $\bar{\mathcal{A}}^{(a)}_\vx = 1$,
      then $\ket{\psi}$ is the exact ground state of the X-cube Hamiltonian (\eqnref{eq:Xcube H}).
  }) on a square membrane with a corner at $\vx=0$.
Such a wavefunction can be obtained by acting on the ground state of the lattice model (\eqnref{eq:Xcube H}) by a product of $\Xh_{\vx,3}$ operators on the membrane.

\begin{figure}
\includegraphics[width=.6\columnwidth]{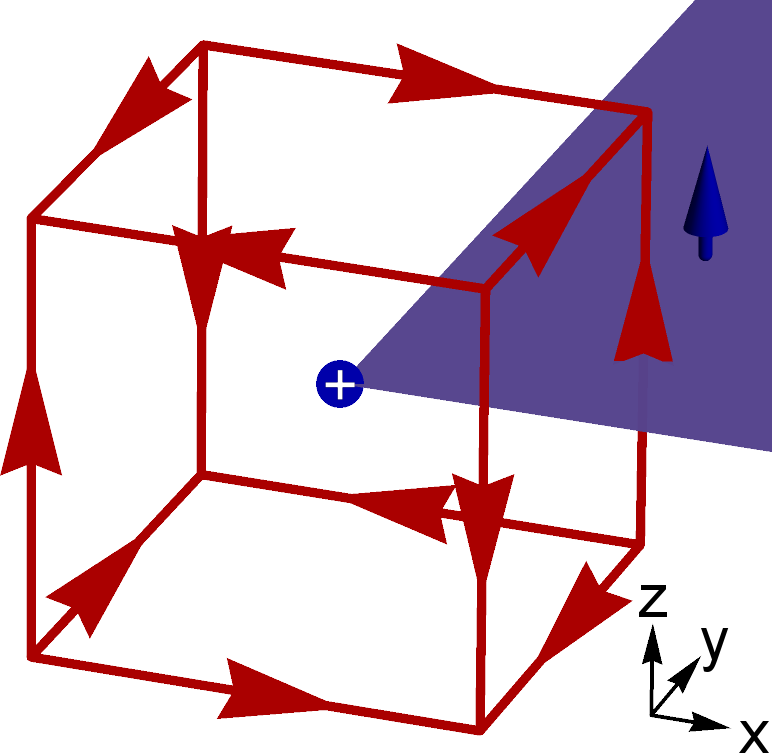}
\caption{
Dimension-1 particles are ``braided'' around a fracton, resulting in a $-2\pi/\N$ phase.
\textbf{(red)} Dimension-1 particle current $J^a$ around the corners of a cube.
\textbf{(blue)} Rectangular membrane where $A_3$ is nonzero.
A single fracton is located at the corner of the membrane, which is inside of the cube.
}\label{fig:dim1Braid}
\end{figure}

In order to obtain a nonzero braiding statistic with the fracton,
  we will consider a dimension-1 particle current $J$ at time $t=0$ around the corners of a cube of length $2\ell$ (\figref{fig:dim1Braid}) which can be written as
\begin{align}
  J^{0;a} &= 0 \nn\\
  J^a &= |\eps^{0abc}| \frac{1}{2} \; \D(t) \sum_{\mu,\nu \is \pm1} \mu\nu\, \Th(\ell + x^a) \Th(\ell - x^a) \label{eq:dim1 J}\\
      &\hspace{3.5cm} \D(x^b - \mu \ell) \D(x^c - \nu \ell) \nn
\end{align}
This current describes a dimension-1 particle which can only move in straight without creating any additional excitations.
In this current configuration, twelve different dimension-1 particles are created so that the edges of a cube are traced out.
(And indeed, these currents satisfy the constraints in \eqnref{eq:Xcube constraint}.)

Although we won't need them, there are a couple nice solutions to \eqref{eq:Xcube EoM'} to describe the current $J$:
\begin{align}
  B_{ab} &= -\frac{2\pi}{\N} \int_{-\infty}^t |\eps_{0abc}| J^c \,\dd t \\
  B_0    &= 0 \nn
\end{align}
which is trivial to integrate since the integral just replaces the $\D(t)$ in $J^a$ by $\Th(t)$.
Another gauge equivalent solution is
\begin{align}
  B_0    &= \frac{2\pi}{\N} \D(t) \prod_a \Th(\ell + x^a) \Th(\ell - x^a) \\
  B_{ab} &= 0 \nn
\end{align}
where $B_0$ is nonzero only inside the cube.

We can now evaluate the action $\int L_\text{X-cube}$ (\eqnref{eq:L Xcube}) for this configuration.
Since the equations of motion for $I^0$ and $J^{0;a}$ are satisfied, the Lagrangian simplifies:
\begin{align}
  L &= \frac{\N}{2\pi} |\eps^{0abc}| \frac{1}{2} B_{ab} \del_0 A_c - A_a J^a - B_{ab} \frac{1}{2} I^{ab} \label{eq:simple Xcube L}\\
    &= - A_a J^a \nn\\
  \int_{t,\vx} L &= -\frac{2\pi}{\N} \nn
\end{align}
The second line results because the first and third terms in the first line are zero since $\del_0 A_a = 0$ (\eqnref{eq:dim1 Z sol}) and $I^{ab} = 0$ (\eqnref{eq:dim1 I}).
Plugging in the expressions for $A_a$ and $J^a$ gives the third line.
The integrand is nonzero only where the red current and blue membrane intersect in \figref{fig:dim1Braid}.
Thus, the presence of a fracton in the cube is detected by ``braiding'' dimension-1 particles around the edges of a cube.

\subsubsection{Fracton ``Braiding''}

As a second example, we show how the presence of a dimension-1 particle can be detected by moving fractons around it. \cite{MaLayers}

The current describing an $x$-axis dimension-1 particle at $\vx=0$ is
\begin{align}
  J^{0;2} &= -J^{0;3} = \DDD(\vx) \\
  J^{0;1} &= J^a = 0 \nn
\end{align}
which has the following field solution:
\begin{align}
  B_{23} &= \frac{2\pi}{\N} \Th(x) \D(y) \D(z) \\
  B_{12} &= B_{13} = B_0 = 0 \nn
\end{align}

\begin{figure}
\begin{minipage}{.47\columnwidth}
\includegraphics[width=\textwidth]{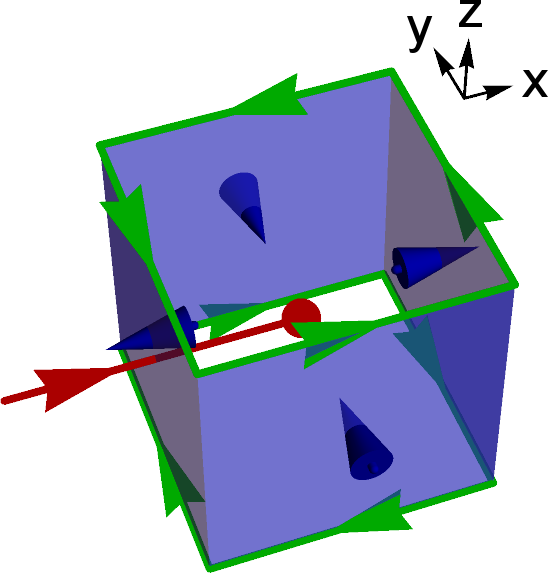}
\textbf{(a)}
\end{minipage}
\hspace{.02\columnwidth}
\begin{minipage}{.47\columnwidth}
\includegraphics[width=\textwidth]{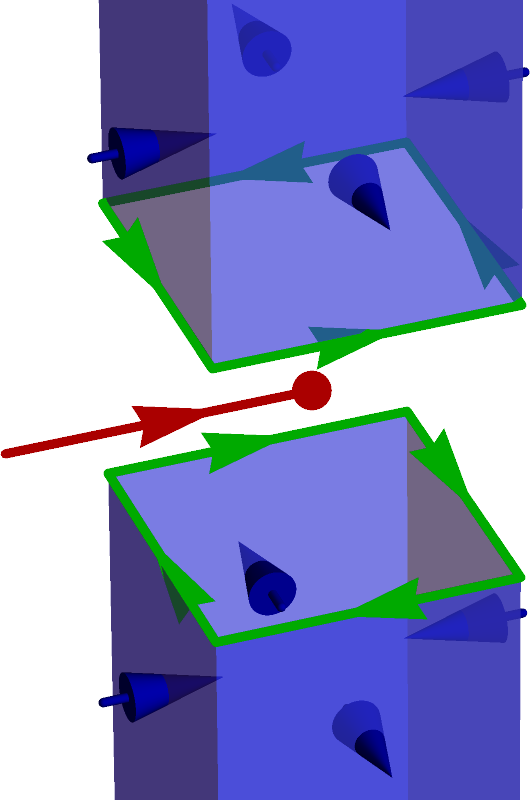}
\textbf{(b)}
\end{minipage}
\caption{
\textbf{(a)}
A pair of oppositely charged fractons are ``braided'' around an $x$-axis dimension-1 particle by exchanging fracton dipoles in the blue shaded region (\figref{fig:fractonCurrent}), resulting in a $+2\pi/\N$ phase.
The fractons are moved parallel to each other;
  the green arrows are used to indicate the sign of the fracton flow $\tI^\A$
\textbf{(b)}
Another fracton current $I$ configuration (blue) which produces the same fracton flow $\tI$ (green).
The blue arrows specify the sign of $B_{ab}$.
However, no phase results from this configuration.
\textbf{(red line)} Region where $B_{23}$ is nonzero.
\textbf{(red point)} An X-axis dimension-1 particle.
\textbf{(blue)} Membrane where the fracton current $I^{ab}$ is nonzero and where $\Xh$ operators are placed in \eqnref{eq:F}.
\textbf{(green)} Fracton flow $\tI$ (\eqnref{eq:fracton flow}).
}\label{fig:fractonBraid}
\end{figure}

To detect the $x$-axis dimension-1 particle within a cube of length $2\ell$,
  we can simultaneously move two oppositely charged fractons around the top and bottom edges of the cube (\figref{fig:fractonBraid}a).
The fractons are capable of moving by exchanging fracton dipoles (see \figref{fig:fractonCurrent}).
However, the current $I$ that describes this process might seem surprising,
  so we'll instead begin with the lattice membrane operator that generates this current and fracton motion (\figref{fig:fractonBraid}a):
\begin{align}
  \hat{\mathcal{F}} &=     \left( \prod_{-\ell < y \leq \ell} \prod_{-\ell < z \leq \ell} \Xh_{-\ell,y,z,1}^\dag \Xh_{+\ell,y,z,1} \right) \label{eq:F}\\
                    &\times \left( \prod_{-\ell < x \leq \ell} \prod_{-\ell < z \leq \ell} \Xh_{x,-\ell,z,2} \Xh^\dag_{x,+\ell,z,2} \right) \nn
\end{align}
The membrane consists of four squares,
  which each create fractons at the corners which cancel out with the fractons generated by neighboring squares.
From the $\Xh$ and $\Zh$ commutation relations (\eqnref{eq:commutator}),
  this operator will rotate the expectation value of $\Zh$.
Using \eqnref{eq:Xcube logZ} and \eqref{eq:redef} to relate $\Zh$ to $A$,
  and the equations of motion for $I^{ab}$ (\eqnref{eq:Xcube EoM'}) to shift $A_c$ via the $\del_0 A_c$ term in \eqnref{eq:Xcube EoM'},
  we find the following fracton current:
\begin{align}
  I^{a3} &= \eps_{0ab3} \, \D(t) \Th(x^a+\ell) \Th(x^a-\ell) \label{eq:fracton current}\\
         &\qquad (\D(x^b + \ell) - \D(x^b - \ell)) \, \Th(z+\ell) \Th(z-\ell) \nn\\
  I^{12} = I^0 &= 0 \nn
\end{align}
$I^{a3}$ is nonzero in the blue regions of \figref{fig:fractonBraid}a.

It may seem surprising that the fracton current $I^{ab}$ (\eqnref{eq:fracton current}) is nonzero on a membrane,
  even though we are trying to describe the movement of two point-like fractons.
However, as explained in \figref{fig:dim1Braid}, this is due to the fact that fractons are created at corners of membrane operators,
  which implies that a membrane operator (or membrane current)
  is required to move a pair of fractons.
This is manifested in the conservation law (copied below) by the second derivative in the second term:
\begin{align}
  \del_0 I^0 - \frac{1}{2} \del_a \del_b I^{ab} = 0 \tag{\ref{eq:Xcube constraint}}\nn
\end{align}
Thus, $I^{ab}$ is best regarded as a fracton dipole current.

In order to understand why we must consider a fracton dipole current in the field theory,
  let us define a new quantity $\tI^\A$, which we'll call the fracton flow, in terms of the fracton current $I$:
\begin{align}
  \tI^0 &= I^0 \label{eq:fracton flow}\\
  \tI^a &= - \frac{1}{2} \del_b I^{ab} \nn
\end{align}
Unlike the fracton current ($I$), the fracton flow $\tI^\A$ equations of motion obey the usual current conservation law:
\begin{align}
  \del_\A \tI^\A \EoM 0 \label{eq:tI constraint}
\end{align}
The fracton flow (green in \figref{fig:fractonBraid}) corresponds to the more intuitive notion of net movement of fractons.
However, on a closed manifold the fracton flow does not uniquely specify the fracton current ($I$),
  nor is it sufficient to calculate the resulting phase from a braiding process.
For example, \figref{fig:fractonBraid}b shows a different current configuration,
  which results in the same fracton flow but a different phase factor;
  on a 3D torus, there is no reason to prefer one current configuration over the other.
Thus, the current $I^{ab}$ carries some addition information:
  fractons can only move by exchanging fracton dipoles (\figref{fig:fractonCurrent}), and
  $I^{ab}$ specifies where the dipole current occurs.

Although we won't need them, $I$ (\eqnref{eq:fracton current}) has a couple of nice field configuration solutions:
\begin{align}
  A_a     &= \frac{2\pi}{\N} \int_{-\infty}^t |\eps_{0abc}| \frac{1}{2} I^{bc} \,\dd t \\
  A_{0;a} &= 0 \nn
\end{align}
Another gauge equivalent solution is
\begin{align}
  A_{0;3} &= - \frac{2\pi}{\N} \D(t) \prod_a \Th(\ell + x^a) \Th(\ell - x^a) \\
  A_{0;1} &= A_{0;2} = A_a = 0 \nn
\end{align}
which is nonzero inside the cube.

We can now evaluate the action $\int L_\text{X-cube}$ (\eqnref{eq:L Xcube}) for this configuration.
Making use of the equations of motion for $I^0$ and $J^{0;a}$
  and the fact that $\del_0 B_{ab} = J^c = 0$ for our current configuration, we find:
\begin{align}
  L &= \frac{\N}{2\pi} |\eps^{0abc}| \frac{1}{2} B_{ab} \del_0 A_c - A_a J^a - B_{ab} \frac{1}{2} I^{ab} \\
    &= -B_{ab} \frac{1}{2} I^{ab} \nn\\
  \int_{t,\vx} L &= + \frac{2\pi}{\N} \nn
\end{align}
Thus, the presence of a dimension-1 particle in the cube is detected by moving a pair of fractons around the top and bottom edges of the cube.
A more detailed analysis would show that this motion of fractons actually counts the difference in the number of $x$-axis and $y$-axis dimension-1 particles inside the cube.

\subsection{Minimal Coupling to Matter}
\label{sec:Xcube matter}

In this section we will show how the X-cube Hamiltonian (\eqnref{eq:Xcube H}) and field theory (\eqnref{eq:L Xcube}) can be coupled to matter,
  which is related to the gauging procedures introduced in \refcite{VijayFracton,WilliamsonUngauging}.
We will leave further study of these models to future work.
See \appref{app:BF matter}, for an analogous treatment for toric code and BF theory in 2+1D.

\begin{figure}
\includegraphics[width=.6\columnwidth]{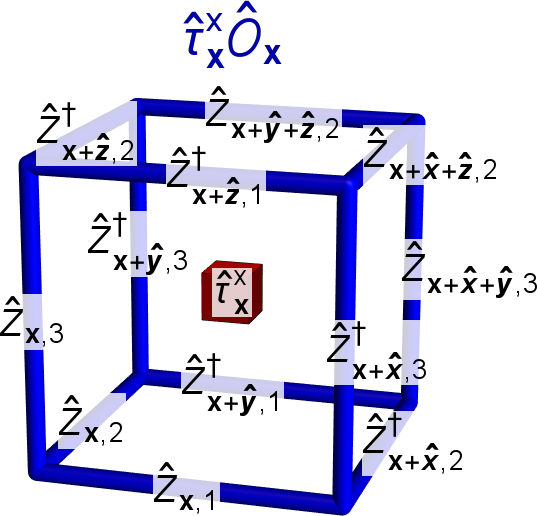}
\includegraphics[width=.38\columnwidth]{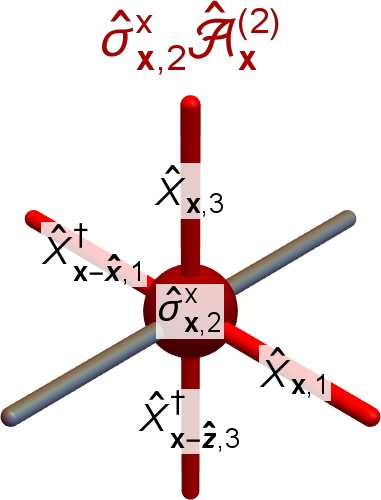} \\
\includegraphics[width=.45\columnwidth]{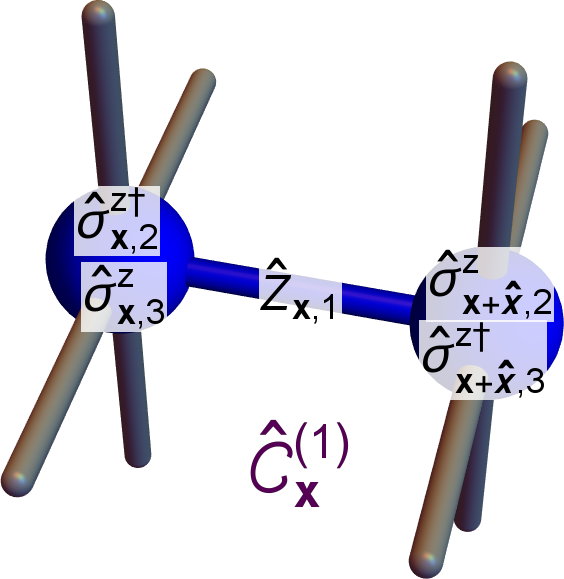}
\includegraphics[width=.4\columnwidth]{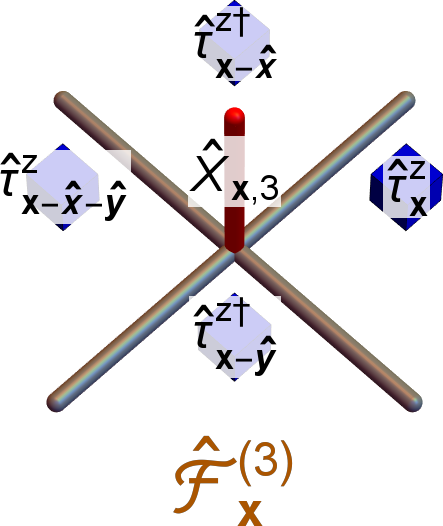}
\caption{
$\hat\tau^x_\vx \Oh_\vx$, $\hat\sigma^x_\vx \Ah^{(a)}_\vx$, $\Ch_\vx^{(a)}$, and $\Fh_\vx^{(a)}$ operators of the X-cube model (\eqnref{eq:Xcube H matter}) after coupling to $\hat\sigma^\mu$ and $\hat\tau^\mu$ matter.
$\hat\sigma^\mu_{\vx,a}$ are centered on the vertices,
  while $\hat\tau^\mu_\vx$ are centered on the cubes.
}\label{fig:Xcube matter}
\end{figure}

In the lattice model, matter can be introduced by introducing $Z_\N$ fracton matter operators $\hat\tau^\mu_\vx$ at the centers of the cubes
  and three dimension-1 matter operators $\hat\sigma^\mu_{\vx,a}$ ($a=1,2,3$) on the sites of the cubic lattice.
The fracton operator $\Oh_\vx$ and and dimension-1 particle operator $\Ah^{(a)}_\vx$ are multiplied by $\hat\tau^x_\vx$ and $\hat\sigma^x_{\vx,a}$, respectively.
We also introduce hopping terms $\Fh_\vx^{(a)}$ and $\Ch_\vx^{(a)}$ for the fracton and dimension-1 matter, respectively (\figref{fig:Xcube matter}).
The Hamiltonian with this matter coupling is
\begin{align}
  \Hh_\text{X-cube}^\text{coupled}
    &= - \sum_\vx \Bigg[\hat\tau^x_\vx \Oh_\vx + \sum_a \Big( \hat\sigma^x_{\vx,a} \Ah_\vx^{(a)} + \Ch_\vx^{(a)} + \Fh_\vx^{(a)} \Big) \nn\\
    &\hspace{1.6cm} + h \sum_a \sigma_{\vx,a}^x + h' \tau_\vx^x \Bigg] + \text{h.c.} \label{eq:Xcube H matter}
\end{align}
$\hat\tau_\vx^x$ and $\hat\sigma_{\vx,a}^x$ are $Z_\N$ fracton and dimension-1 matter number operators,
  and ``h.c.'' denotes the addition of the Hermitian conjugate of the preceding operators.

If $h$ and/or $h'$ is small, then the $A$ and/or $B$ gauge fields are ``Higgsed'',
  and $\Hh_\text{X-cube}^\text{coupled}$ is in a trivial phase with no topological order.
This occurs because the Wilson and 't Hooft loop operators (\figref{fig:fractonLoops}), which describe the ground state degeneracy,
  don't commute with the $\Fh$ and $\Ch$ operators, respectively.
When $h$ and $h'$ are large, the matter has a large mass gap and has no effect on the phase.

The fracton matter hopping operator ($\Fh$) hops fracton matter (i.e. excitations of $\hat\tau_\vx^x$) with the same mobility constraints as the fracton excitations in the original X-cube model (\eqnref{eq:Xcube H}).
For example, it can create four fractons from the vacuum,
  or it can move a fracton dipole along a plane as fracton dipoles are dimension-2 particles.
Similarly, the dimension-1 matter hopping operator ($\Ch$) hops dimension-1 matter (i.e. excitations of $\hat\sigma_{\vx,a}^x$) along straight lines.
Thus, $\hat\tau_\vx^x$ and $\hat\sigma_{\vx,a}^x$ are analogous to the $\Oh$ and $\Ah^{(a)}$ operators, respectively,
  in the X-cube model without explicit matter coupling (\eqnref{eq:Xcube H}).
In the X-cube model, the mobility constraints of the fracton and dimension-1 particle excitations was robust;
  i.e. stable to arbitrary local perturbations.
If we were to consider the matter content of $\Hh_\text{X-cube}^\text{coupled}$ (\eqnref{eq:Xcube H matter}) in the absence of the gauge fields $A$ and $B$,
  then the mobility constraints of the matter would instead be enforced by subdimensional symmetries.

We can describe the same physics in the field theory by introducing $2\pi$-periodic (i.e., $2\pi$ vortices are allowed) matter fields $\theta_a$ and $\phi$.
The simplest way to systematically construct a gauge invariant Lagrangian is to first define currents $j$ and $i$ (not to be confused with $\J$ and $\I$ in \eqnref{eq:L Xcube'})
  equal to $-v^2 A$ and $-w^2 B$, respectively, and then apply a gauge transformation (\eqnref{eq:Xcube gauge} and \eqref{eq:Xcube gauge'}) with $\zeta_a = -\theta_a$ and $\chi = -\phi$:
\begin{align}
  j_{0;a} &= v^2 (\del_0 \theta_a - A_{0;a}) \\
  j_a     &= v^2 \left( \sum_{bc} \eps^{0abc} \del_a \theta_c + A_a \right) \nn\\
  i_0     &= w^2 (\del_0 \phi - B_0) \nn\\
  i_{ab}  &= w^2 (\del_a \del_b \phi - B_{ab}) \nn
\end{align}
where $i_{ab}$ is off-diagonal and symmetric,
  and $v$ and $w$ are coupling constants.
We can now define a Lagrangian for the matter fields:
\begin{align}
  L_\text{X-cube}^\text{coupled} &= L_\text{X-cube} + \sum_a \frac{1}{2v^2} [(j_{0;a})^2 - (j_a)^2] \label{eq:Xcube matter}\\
    &\quad + \frac{1}{2w^2} \left[(i_0)^2 - \sum_{a\neq b} \frac{1}{2} (i_{ab})^2\right] \nn\\
  \sum_a \theta_a &= 0 \label{eq:theta constraint}
\end{align}
where we have imposed a local constraint on $\theta_a$,
  analogous to the local constraint placed on $A_{0;a}$ (\eqnref{eq:L Xcube Hspace}).
(Without the constraint, $\theta$ would have a trivial local symmetry $\theta_a(t,\vx) \rightarrow \tilde\zeta(t,\vx)$.)
\footnote{Note, in this work we are writing all Lagrangians in real time for consistency;
  in imaginary time two signs will flip in \eqnref{eq:Xcube matter} so that $L_\text{X-cube}^\text{coupled}$ is positive definite.}

The advantage of this construction is that the Lagrangian is gauge invariant as long as $\theta$ and $\phi$ transform as
\begin{align}
  \theta_a(t,\vx) &\rightarrow \theta_a(t,\vx) + \zeta_a(t,\vx) \label{eq:Xcube matter gauge}\\
  \phi(t,\vx) &\rightarrow \phi(t,\vx) + \chi(t,\vx) \nn
\end{align}
This construction also guarantees that the equations of motion for $\theta$ and $\phi$
  imply that the matter currents $j$ and $i$ obey the mobility (or generalized charge conservation) constraints in \eqnref{eq:Xcube constraint}.

Note that before $\phi$ is coupled to the gauge field $B$ (e.g. set $B=0$ in $L_\text{X-cube}^\text{coupled}$),
  $\phi$ has a subdimensional symmetry
\begin{equation}
  \phi(t,\vx) \rightarrow \phi(t,\vx) + \sum_a \tilde\chi_a(x^a) \label{eq:phi subdim}
\end{equation}
  where $\tilde\chi_a$ is time independent and depends on only a single spatial coordinate $x^a$.
$\phi$ is analogous to $\hat\tau$ (in \eqnref{eq:Xcube H matter}) and describes the fracton matter
  where the subdimensional symmetry (\eqnref{eq:phi subdim}) protects the fracton mobility constraints.
Gauging $\phi$ then promotes the subdimensional symmetry to a local symmetry (\eqnref{eq:Xcube matter gauge}).

Similarly, before $\theta$ is coupled to the gauge field $A$,
  $\theta$ also has a subdimensional symmetry
\begin{align}
  \theta_a(t,\vx) &\rightarrow \theta_a(t,\vx) + \sum_b \tilde\zeta_b(x^b) - 3 \tilde\zeta_a(x^a) \label{eq:theta subdim}
\end{align}
where $\tilde\zeta_a$ only depends on $x^a$.
$\theta$ is analogous to $\hat\sigma$ (in \eqnref{eq:Xcube H matter}) and describes the dimension-1 matter
  where the subdimensional symmetry (\eqnref{eq:theta subdim}) protects the dimension-1 particle mobility constraints.
Gauging $\theta$ promotes the subdimensional symmetry to a local symmetry (\eqnref{eq:Xcube matter gauge}).

Alternatively, we can couple the X-cube field theory to a complex-valued scalar field $\Phi$,
   which describes the fracton matter, by introducing amplitude components to $\phi$;
   i.e. $\Phi = |\Phi| e^{\ii \phi}$.
(A treatment of the dimension-1 matter ($\theta_a$) is omitted since it is more involved due to the constraint \eqnref{eq:theta constraint}.)
The phase of $\Phi$ transforms accordingly under gauge (\eqnref{eq:Xcube matter gauge}) and subdimensional symmetry (\eqnref{eq:phi subdim}) transformations.
The matter currents generalize to\footnote{%
  Previous versions of this work were missing the $\del_a \Phi \del_b \Phi$ term,
    which is necessary for gauge invariance.
    This term was recently introduced by Michael Pretko in \refcite{PretkoGauge} for the $U(1)$ scalar charge tensor gauge theory.}
\begin{align}
  \tilde i_0     &= w (\del_0 - \ii\, B_0) \Phi \label{eq:amplitude currents}\\
  \tilde i_{ab}  &= \Phi (\del_a \del_b + B_{ab}) \Phi - \del_a \Phi \del_b \Phi \nn
\end{align}
The Lagrangian takes a similar form
\begin{align}
  L_\text{X-cube}^{\text{coupled}'} &= L_\text{X-cube} + \frac{1}{2w^2} \left[|\tilde i_0|^2 - \sum_{a\neq b} \frac{1}{2} |\tilde i_{ab}|^2\right] \label{eq:Xcube matter'}\\
    &\quad - \mu\, (|\Phi|^2 - w^2)^2 \nn
\end{align}
When $\mu$ is large, $\Phi \approx w\, e^{\ii \phi}$ and we reproduce \eqnref{eq:Xcube matter}
  (aside from the omission of the dimension-1 matter ($\theta_a$)).


\subsection{Subconformal Invariance}
\label{sec:Xcube invariance}

The Lagrangians for BF theory in 2+1D (\eqnref{eq:L BF}) and 3+1D (\eqnref{eq:L 3+1D BF}) are topologically invariant (\appref{app:BF invariance}).
(That is, they are invariant under smooth spacetime transformations, which preserve the topology of the spacetime manifold.)
As a result, the ground state degeneracy only depends on the topology of the spatial manifold,
  and the braiding statistics only depend on the topology of the paths of the particle (and string) excitations.
However, the ground state degeneracy of the X-cube model depends on the system size,
  and the dimension-1 particles aren't even allowed to move in all directions.
Thus, a field theory of the X-cube model can not be topologically invariant since even rotation symmetry is broken.
Nevertheless, in this section we will show that our X-cube field theory (\eqnref{eq:L Xcube}) is invariant under a certain subgroup of the conformal group of spacetime transformations,
  which we will refer to as the subconformal group.
(The conformal group is the group of spacetime transformations that preserves angles.)

Specifically, the X-cube field theory is invariant under the following spacetime transformation:
\begin{align}
  t &\rightarrow \tT(t) \label{eq:subconformal transformation}\\
  x &\rightarrow \tX(x) \nn\\
  y &\rightarrow \tY(y) \nn\\
  z &\rightarrow \tZ(z) \nn
\end{align}
where $\tT(t)$, $\tX(x)$, $\tY(y)$, and $\tZ(z)$ are smooth and monotonic.
Note that each spacetime component transforms independently of every other component.
This is necessary due to the physics of dimension-1 particles which can only move in straight lines:
  the spacetime transformation is not allowed to bend or distort the spatial geometry.
The fields transform as usual (i.e. the same as in BF theory (\appref{app:BF invariance}) given the transformation \eqnref{eq:subconformal transformation}), except for the fact that indices following a semicolon do not transform:
\begin{align}
  A_{0;a}(\R) &\rightarrow \tilde{A}_{0;a}(\R) = \frac{\dd \tT}{\dd t} A_{0;a}(\tRR) \label{eq:subconformal}\\
  A_a(\R)     &\rightarrow \tilde{A}_a(\R)     = \frac{\dd \tX^a}{\dd x^a} A_a(\tRR) \nn\\
  B_0(\R)     &\rightarrow \tilde{B}_0(\R)     = \frac{\dd \tT}{\dd t} B_0(\tRR) \nn\\
  B_{ab}(\R)  &\rightarrow \tilde{B}_{ab}(\R)  = \frac{\dd \tX^a}{\dd x^a} \frac{\dd \tX^b}{\dd x^b} B_{ab}(\tRR) \nn\\
  J^{0;a}(\R) &\rightarrow \tilde{J}^{0;a}(\R) = \frac{\dd \tX}{\dd x} \frac{\dd \tY}{\dd y} \frac{\dd \tZ}{\dd z} J^{0;a}(\tRR) \nn\\
  J^a(\R)     &\rightarrow \tilde{J}^a(\R)     = |\eps^{0abc}| \frac{1}{2} \frac{\dd \tT}{\dd t} \frac{\dd \tX^b}{\dd x^b} \frac{\dd \tX^c}{\dd x^c} J^a(\tRR) \nn\\
  I^0(\R)     &\rightarrow \tilde{I}^a(\R)     = \frac{\dd \tX}{\dd x} \frac{\dd \tY}{\dd y} \frac{\dd \tZ}{\dd z} I^0(\tRR) \nn\\
  I^{ab}(\R)  &\rightarrow \tilde{I}^{ab}(\R)  = |\eps^{0abc}| \frac{\dd \tT}{\dd t} \frac{\dd \tX^c}{\dd x^c} I^{ab}{a}(\tRR) \nn
\end{align}
It is then simple to see that the action $\int L_\text{X-cube}$ (\eqnref{eq:L Xcube}) is invariant under the above transformation.

\subsection{Robust Degeneracy}
\label{sec:Xcube degen}

The ground state degeneracy of the X-cube model is robust in the sense that
  generic local perturbations to the Hamiltonian or Lagrangian do not split the degeneracy.
(More precisely, on a finite system the degeneracy splitting is exponentially small in system size.)
However, the degeneracy is not topological in the sense that the degeneracy depends on more than the topology of the spatial manifold:
  a fact that we will explore further in \secref{sec:curvature degeneracy}.
On a torus, the degeneracy is exponentially large with the lengths of the torus.
In this section, we explain how to calculate the ground state degeneracy of the X-cube model and its field theory.
We also express the degeneracy in terms of an integral (\eqnref{eq:Xcube degen integral}), which is invariant under the subconformal transformation (\secref{sec:Xcube invariance}).

\subsubsection{Lattice Model Degeneracy on a 3D Torus}

\begin{figure}
\includegraphics[width=.6\columnwidth]{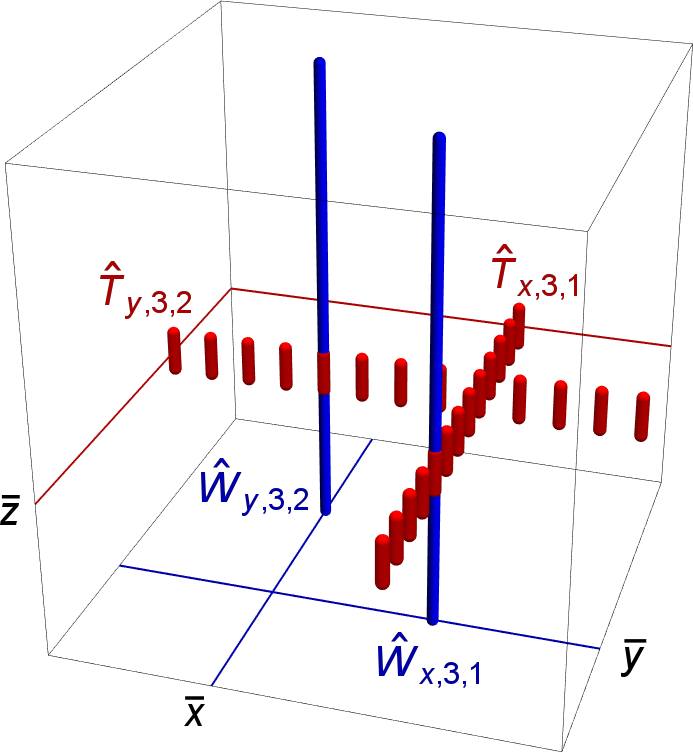}
\caption{
Examples of loop operators given in \eqnref{eq:Xcube loop operators} on a periodic cubic lattice.
A $\hat{W}$ operator moves a dimension-1 particle around the torus,
  while the $\hat{T}$ operator moves a fracton dipole.
$\Zh$ and $\Xh$ operators are placed on the blue and red colored links, respectively.
The loop operators are parameterized by a coordinate.
For example, for each $x$ there is a $\hat{W}_{x,3,1}$ operator;
  changing $x$ moves the loop operator along the blue line at the bottom.
Modulo the redundancy in \eqnref{eq:loop redundancy},
  these operators are the nonlocal qubits which act on the degenerate ground state manifold of the X-cube model (\eqnref{eq:Xcube H}).
}\label{fig:fractonLoops}
\end{figure}

First, we will explain how the degeneracy of the X-cube lattice model can be understood in terms of nonlocal qubit operators on an $L_1 \times L_2 \times L_3$.
(Other derivations of the degeneracy are given in \cite{VijayXCube,VijayLayer,MaLayers}.)
Similar to toric code, the X-cube model also has noncontractible loop operators (\figref{fig:fractonLoops}):
\begin{align}
              &\hat{W}_{x,3,1} = \prod_z \Zh_{x  ,\by,z  ,3}
  \quad,\quad  \hat{T}_{x,3,1} = \prod_y \Xh_{x  ,y  ,\bz,3} \label{eq:Xcube loop operators}\\
              &\hat{W}_{y,3,2} = \prod_z \Zh_{\bx,y  ,z  ,3}
  \quad,\quad \hat{T}_{y,3,2} = \prod_x \Xh_{x  ,y  ,\bz,3} \nn\\
  &\hat{W}_{y,1,2}, \hat{W}_{z,1,3}, \hat{W}_{x,2,1}, \hat{W}_{z,2,3} \text{ and corresponding} \nn\\
  &\hat{T} \text{ operators are similarly defined.} \nn
\end{align}
where $\bx$, $\by$, and $\bz$ are arbitrary constants.
$\hat{W}$ and $\hat{T}$ correspond to moving a dimension-1 particle and a fracton dipole around the torus, respectively.
These operators are observables capable of distinguishing the degenerate ground states.

The above operators have the following nontrivial commutation relations:
\begin{align}
  \hat{T}_{x,3,1} \hat{W}_{x',3,1} &= \omega^{\D_{x,x' }} \hat{W}_{x',3,1} \hat{T}_{x,3,1}    \\
  \hat{T}_{y,3,2} \hat{W}_{y',3,2} &= \omega^{\D_{y,y' }} \hat{W}_{y',3,2} \hat{T}_{y,3,2} \nn\\
  \hat{T}_{y,3,2} \hat{W}_{x ,3,1} &= \omega^{\D_{y,\by}} \hat{W}_{x ,3,1} \hat{T}_{y,3,2} \nn\\
  \hat{T}_{x,3,1} \hat{W}_{y ,3,2} &= \omega^{\D_{x,\bx}} \hat{W}_{y ,3,2} \hat{T}_{x,3,1} \nn\\
  \omega &= e^{2\pi \ii/\N} \tag{\ref{eq:commutator}} \nn
\end{align}
and similar for $\hat{W}_{y,1,2}$, $\hat{W}_{z,1,3}$, $\hat{W}_{x,2,1}$, $\hat{W}_{z,2,3}$,
  and the corresponding $\hat{T}$ operators.
Thus, $\hat{W}$ and $\hat{T}$ are conjugate operators,
  with the exception of some redundancy which resulted in the third and fourth equations above.
The redundancy is:
\begin{align}
          \hat{W}_{\bx,3,1} &=         \hat{W}_{\by,3,2} \label{eq:loop redundancy}\\
  \prod_x \hat{T}_{x  ,3,1} &= \prod_y \hat{T}_{y  ,3,2} \nn
\end{align}
and similar for $\hat{W}_{y,1,2}$, etc.
We can define new loop operators without this redundancy:
\begin{align}
  \hat{W}'_{x,3,1}  &= \hat{W}_{x,3,1} \label{eq:Xcube loop operators''}\\
  \hat{W}'_{y,3,2}  &= \hat{W}_{y,3,2} \hat{W}_{\by,3,2}^\dag \nn\\
  \text{and ignore }& \hat{W}'_{\by,3,2} \text{ and } \hat{T}_{\by,3,2} \nn
\end{align}
and similar for $\hat{W}_{y,1,2}$, etc.
If we use $\hat{W}'$ instead of $\hat{W}$, and ignore $\hat{W}'_{\by,3,2}$ and $\hat{T}_{\by,3,2}$,
  then the only nontrivial commutation relations are
\begin{align}
  \hat{T}_{x,3,1} \hat{W}'_{x',3,1} &= \omega^{\D_{x,x' }} \hat{W}'_{x',3,1} \hat{T}_{x,3,1}    \\
  \hat{T}_{y,3,2} \hat{W}'_{y',3,2} &= \omega^{\D_{y,y' }} \hat{W}'_{y',3,2} \hat{T}_{y,3,2} \nn
\end{align}
and similar for $\hat{W}_{y,1,2}$, etc.
Therefore, we have $(L_1 + L_2 - 1) + (L_2 + L_3 - 1) + (L_3 + L_1 - 1) = 2L_1 + 2L_2 + 2L_3 - 3$ independent pairs of conjugate $Z_\N$ operators that commute with the X-cube Hamiltonian (\eqnref{eq:Xcube H}).
The degeneracy is therefore
\begin{align}
  \text{degen} &= \N^{2L_1 + 2L_2 + 2L_3 - 3} \label{eq:Xcube degen}
\end{align}

\subsubsection{Field Theory Degeneracy on a Torus}

Now we want to calculate the ground state degeneracy of the X-cube field theory on an $l^1 \times l^2 \times l^3$ torus.
(The superscripts in $l^a$ are spatial indices, not exponents.)
However, now the lengths $l^a$ have units of length and are no longer integers.

To understand how to deal with this issue, let us first consider the example of a stack of $L_3$ layers of $Z_\N$ toric codes.
The degeneracy on an $L_1 \times L_2 \times L_3$ torus is $\N^{2L_3}$
  \footnote{This degeneracy is stable to perturbations since on a finite lattice the degeneracy is only lifted at order $\sim \min(L_x,L_y)$
              and therefore the energy splitting is exponentially small with system size.}.
A natural field theory for this model is the 2+1D BF theory (\eqnref{eq:L BF}) with an extra dimension in the $z$ direction:
\begin{align}
  L_\text{BF stack} &= \frac{\N}{2\pi} \sum_{\A\B\C=0,1,2} \epsilon^{\A\B\C} B_\A(t,x,y,z) \del_\B A_\C(t,x,y,z)
\end{align}
Note that there is no $z$ derivative $\del_3$;
  each layer is decoupled.
Since the $z$ direction is continuous in the field theory,
  the ground state degeneracy on a torus appears to be infinite.
However, if impose a UV cutoff length $a$,
  then the degeneracy is finite:
\begin{align}
  \text{degen} \sim \N^{2l^3/a}
\end{align}
for a torus of length $l^1 \times l^2 \times l^3$.
This is the degeneracy because the effective number of layers is $L_3 \sim l^3/a$ and the lattice model degeneracy is $\N^{2L_3}$.
(Recall that $l^3$ is not $l$ cubed; it's the $z$ component of $\textit{\textbf{l}}$.)
Thus, the UV cutoff allows us to calculate a finite degeneracy.

With this in mind, we can now understand the degeneracy of the X-cube field theory (\eqnref{eq:L Xcube}) on an $l^1 \times l^2 \times l^3$ torus.
Integrating over $A_{0;a}$ and $B_0$ enforces a zero fracton and dimension-1 particle constraint:
  $I^0 = J^{0;a} = 0$ (where $I^0$ and $J^{0;a}$ are given by their equations of motion \eqnref{eq:Xcube EoM'}).
Modulo gauge redundancy (\eqnref{eq:Xcube gauge}), on an $l^1 \times l^2 \times l^3$ torus the solutions to these constraints can be written as
\begin{align}
  A_a  (t,\vx)  &= |\eps^{0abc}| \D(x^a-\bx^a) q_{;ab}(t,x^b) \label{eq:Xcube sol}\\
  B_{ab}(t,\vx) &= |\eps_{0abc}| \D(x^b-\bx^b) \bigg[ p^{;ca}(t,x^a) \nn\\
                & - \frac{1-\eps^{0abc}}{2} \D(x^a - \bx^a) \int_{\tX^a} p^{;ca}(t,\tX^a) \bigg] + (a \leftrightarrow b) \nn\\
  q_{;ab}(t,\bx^b) &= p^{;ab}(t,\bx^b) = 0 \quad\text{if}\quad \eps^{0abc} = -1 \label{eq:Xcube sol constraints}
\end{align}
where $q_{;ab}(t,x^b)$ and $p^{;ab}(t,x^b)$ describe the topological contribution.
The second two lines are accounting for the redundancy described in \eqnref{eq:loop redundancy}.
Without this correction, the action (\eqnref{eq:Xcube degen action}) would contain additional unwanted terms.
$p^{;ab}$ and $q_{;ab}$ are closely connected to $\hat{W}'$ and $\hat{T}$ in \eqnref{eq:Xcube loop operators''} and \eqref{eq:Xcube loop operators}:
\begin{align}
  \hat{W}'_{x^b,a,b}(t) &\sim e^{\ii\, p^{;ab}(t,x^b)} &,\quad
  \hat{T} _{x^b,a,b}(t) &\sim e^{\ii\, q_{;ab}(t,x^b)}
\end{align}
For example:
\begin{align}
  \hat{W}'_{x,3,1}(t) &=    \prod_y \Xh_{x,y,\bz,3} \nn\\
                      &\sim \exp(\ii \int_y B_{12}(t,x,y,\bz)) \nn\\
                      &=    e^{\ii\, p^{;31}(t,x)} \nn
\end{align}

In a different gauge, $A_a$ and $B_{ab}$ can also be written as
\begin{align}
  A_a  (t,\vx) &= |\eps^{0abc}| q_{;ab}(t,x^b)/l^a \label{eq:Xcube sol'}\\
  B_{ab}(t,\vx) &= |\eps_{0abc}| \bigg[ p^{;ca}(t,x^a) \nn\\
                &\qquad - \frac{1-\eps^{0abc}}{2} \int_{\tX^a} p^{;ca}(t,\tX^a) / l^a \bigg] \nn\\
  \int_{x^b} q_{;ab}(t,x^b) &= \int_{x^b} p^{;ab}(t,x^b) = 0 \quad\text{if}\quad \eps^{0abc} = -1 \label{eq:Xcube sol constraints'}
\end{align}
This choice makes $A_a$ and $B_{ab}$ more smooth by removing the delta functions,
  but is less closely connected to $\hat{W}'$ and $\hat{T}$.

If we insert either \eqnref{eq:Xcube sol} or \eqref{eq:Xcube sol'} into the action $\int L_\text{X-cube}$ (\eqnref{eq:L Xcube}), we find
\begin{align}
  \int_{t,\vx} L &= \frac{\N}{2\pi} |\eps^{0abc}| \int_{t,x^b} p^{;ab}(t,x^b) \del_0 q_{;ab}(t,x^b) \label{eq:Xcube degen action}
\end{align}
With a UV cutoff $a$, this action describes $2l^1/a + 2l^2/a + 2l^3/a - 3$ qubits
  ($\hat{W}'$ and $\hat{T}$).
The $-3$ comes from the three pairs of constraints in \eqnref{eq:Xcube sol constraints} or \eqref{eq:Xcube sol constraints'}.
Thus, the field theory has the following ground state degeneracy
\begin{align}
  \text{degen} \sim \N^{2l^1/a + 2l^2/a + 2l^3/a - 3}
\end{align}
which is consistent with the X-cube lattice model degeneracy (\eqnref{eq:Xcube degen}).

Note that this expression for the degeneracy can be made subconformally invariant (\secref{sec:Xcube invariance}) if we generalize it to an integral:
\begin{align}
  \log_{\N}\text{degen} &\sim -3 + \sum_{bcd} \int_\vx \frac{|\eps^{0bcd}|}{l^b l^c a^d(x^d)} \label{eq:Xcube degen integral}
\end{align}
where we now allow the cutoff $a^d(x^d)$ to be position and direction dependent.
This integral is only appropriate for a flat $l^1 \times l^2 \times l^3$ torus.
A more general expression (which we leave to future work) is necessary for curved spaces (\secref{sec:curvature degeneracy}).
Under the subconformal transformation, the cutoff transforms as (using the notation of \secref{sec:Xcube invariance}):
\begin{align}
  a^b(x^b) \rightarrow \tilde{a}^b(x^b) = \frac{\dd x^b}{\dd \tX^b} a^b(\tX^b(x^b))
\end{align}

\section{Robust Degeneracy via Lattice Curvature}
\label{sec:curvature degeneracy}

For toric code and BF theory on a manifold without a boundary,
  the ground state degeneracy only depends on the topology of the manifold.
However, as reviewed in the previous section,
  the degeneracy of the X-cube model on a torus depends on the size of the torus,
  which is a geometric (not topological) property.
Since the topology of the spatial manifold is insufficient to determine the ground state degeneracy,
  it is plausible that a manifold with nontrivial topology may not even be necessary for a ground state degeneracy.
Indeed, in this section we will demonstrate that the X-cube model can have a stable ground state degeneracy on a manifold with trivial topology!
We will focus on lattice models in this section and leave a field theory description to future work.
For simplicity, we take $\N=2$ here, so that $\Zh = \Zh^\dag$ and $\Xh = \Xh^\dag$,
  which makes it easier to generalize the X-cube model to more complicated lattices.

\begin{figure}
\begin{minipage}{.4\columnwidth}
\includegraphics[width=\textwidth]{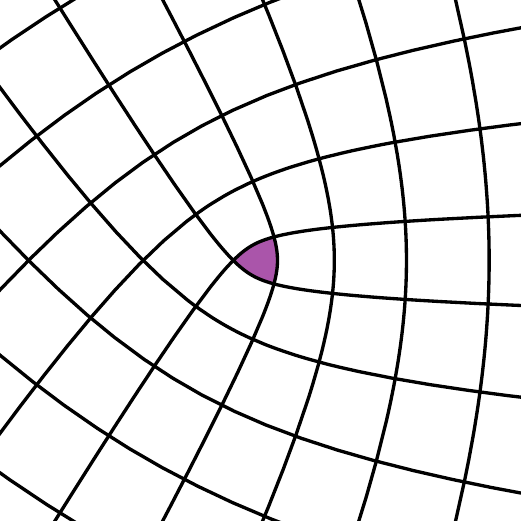}
\textbf{(a)} -90\degree\ defect
\end{minipage}
\hspace{.08\columnwidth}
\begin{minipage}{.4\columnwidth}
\includegraphics[width=\textwidth]{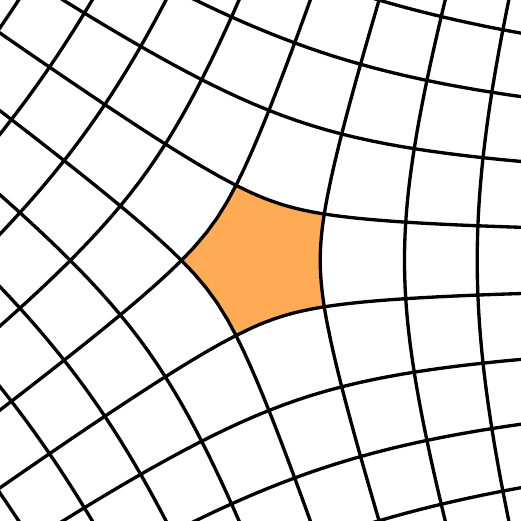}
\textbf{(b)} +90\degree\ defect
\end{minipage}
\vspace*{.4cm}\\
\begin{minipage}{.55\columnwidth}
\includegraphics[width=\textwidth]{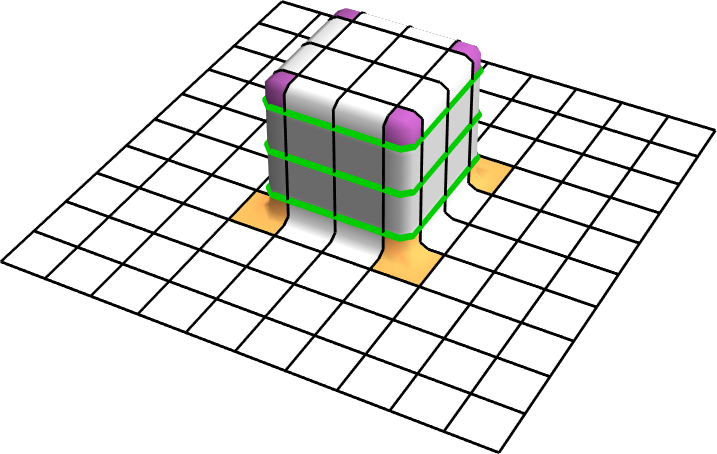}
\textbf{(c)}
\end{minipage}
\hspace{.01\columnwidth}
\begin{minipage}{.4\columnwidth}
\includegraphics[width=\textwidth]{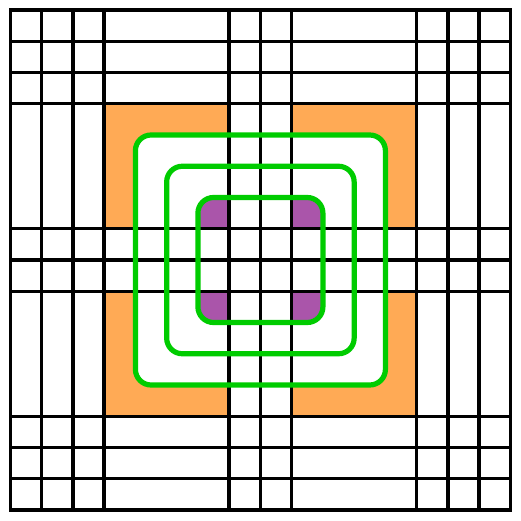}
\textbf{(d)}
\end{minipage}
\caption{
Examples of 2D lattices with curvature defects.
\textbf{(a,b)} $\pm90\degree$ square lattice defects.
\textbf{(c)}
A square lattice with four -90\degree\ defects (purple) and four +90\degree\ defects (orange).
The 2D lattice is embedded in 3D space
  where the 2D surface takes the shape of an oversized tablecloth that is placed over a square table.
The resulting curvature of the lattice leads to the creation of ``straight loops'' around the square:
  i.e. a geodesic loop without any kinks.
These loops are colored green.
Straight loops are important because they can be traversed by a dimension-1 particle
  (via loop operators as in \figref{fig:fractonLoops}).
\textbf{(d)}
The same lattice, but embedded in 2D space by stretching some of the plaquettes and drawing new (green) links.
}\label{fig:squareTable}
\end{figure}

We will consider examples of a curved lattices with angular defects (\figref{fig:squareTable}a-b).
\figref{fig:squareTable}c-d shows an instructive example of a 2D lattice with angular defects with an important property:
  there are ``straight loops'' (green in figure).
These loops are straight in the sense that they can be traversed by dimension-1 particles.

\begin{figure}
\begin{minipage}{.54\columnwidth}
\includegraphics[width=\textwidth]{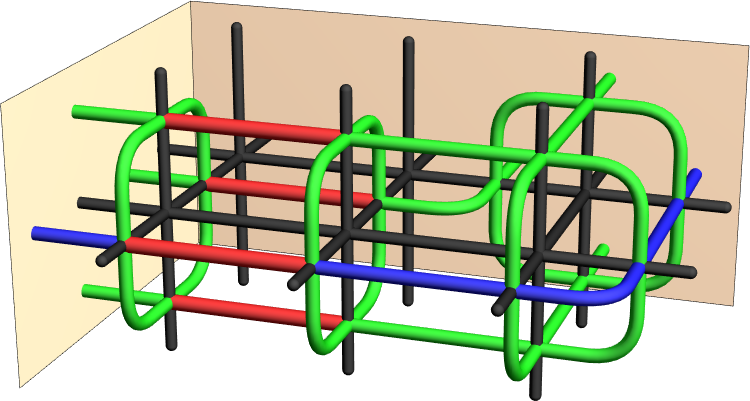}
\textbf{(a)}
\end{minipage}
\begin{minipage}{.44\columnwidth}
\includegraphics[width=\textwidth]{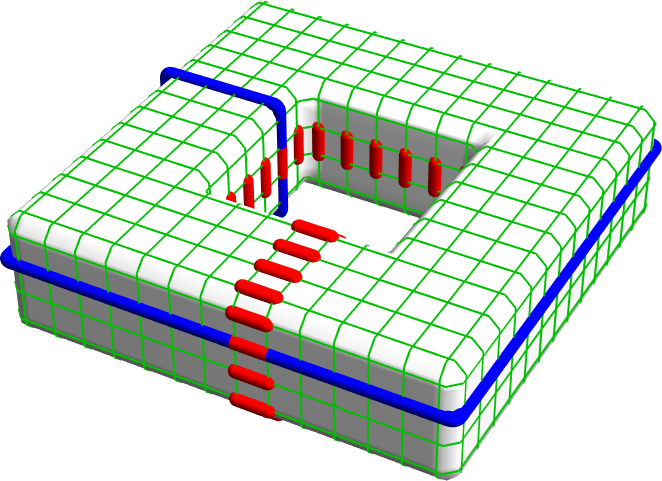}
\textbf{(b)}
\end{minipage}
\caption{
A cubic lattice with many angular defects.
\textbf{(a)}
Similar to \figref{fig:squareTable}d:
We start with a cubic lattice (black) and then add the colored links.
For clarity, we only show the $y<0<x$ quadrant;
  the other four quadrants are obtained by reflecting about the two light-orange colored planes.
The colored links thus take the shape of a torus.
\textbf{(b)}
A larger example: we start with a cubic lattice (not shown for clarity) and then add the green links.
This time, the green links form a larger torus.
If this geometry were embedded in 4D space,
  it would take the shape of a 3D plane with a torus raised into an orthogonal 4th dimension
  (similar to \figref{fig:squareTable}c, which took the shape of a 2D plane with a square raised into the 3rd dimension).
When $\N=2$, the above lattices have a ground state degeneracy $2^2$ due to the two pairs of red and blue loop operators
  (analogous to the operators in \figref{fig:fractonLoops}).
Note that the topology of the above lattice is trivial:
  its topology is the same as the original cubic lattice.
The degeneracy results from the geometry,
  not the topology!
(Although it appears similar, the physics on the green lattice is not the same as toric code on a torus; e.g. toric code lacks dimension-1 particles.)
}\label{fig:torusTable}
\end{figure}

In \figref{fig:torusTable} we now show the 3D lattice example of interest:
  a cubic lattice with angular defects around the edges of a rectangular torus.
Importantly, the lattice has straight loops (blue in figure) which can be traversed by dimension-1 particles.
When $\N=2$, the lattice has ground state degeneracy $2^2$ due to the two pairs of loop operators (red and blue in the figure),
  which are analogous to the operators in \figref{fig:fractonLoops}.
Since the topology of the lattice is trivial
  (its topology is the same as the original cubic lattice),
  the degeneracy is a result of the \emph{geometry} of the lattice,
  not the topology!
The degeneracy is robust in the sense that if perturbations are added,
  the energy splitting is exponentially small in the size of the torus-shaped (green in figure) curved region
\footnote{This can be shown using degenerate perturbation theory,
  which will have to generate a large loop operator around the torus at a large perturbative order.}.
The degeneracy can also be made exponentially large by adding more layers.
\figref{fig:torusTable} only adds a single layer of links (green in figure);
  if $\ell$ layers are added (e.g. \figref{fig:squareTable}c-d has 3 layers),
  then the degeneracy is $2^{2\ell}$ (when $\N=2$).

Some of the cubes in \figref{fig:torusTable} have been split into two 3-cells.
(Links, plaquettes, and cubes are examples of 1, 2, and 3--cells.)
Thus, for each of these cubes, the original fracton operator $\Oh$ (\figref{fig:Xcube}) in the Hamiltonian $\Hh_\text{X-cube}$ (\eqnref{eq:Xcube H}),
  must be replaced by two new operators (one for each new 3-cell),
  which are each a product of $\Zh$ operators on the links on the edge of the 3-cell.
Also, vertex operators $\Ah^{(a)}$ will need to be added to $\Hh_\text{X-cube}$ at the new vertices (where the black and green lines intersect in \figref{fig:torusTable}a).
(Recall that we're only considering $\N=2$ so that $\Zh = \Zh^\dag$ and $\Xh = \Xh^\dag$, and thus we don't need to worry about which operators require complex conjugation.)

\section{Conclusion}

In this work we have exemplified a generic method to derive a quantum field theory from either certain exactly solvable lattice models
  or from a charge density (e.g. \eqnref{eq:toric EoM}, \eqref{eq:Xcube EoM}, or \eqref{eq:3+1D EoM}).
We used the method to derive a quantum field theory (\eqnref{eq:L Xcube}) for the X-cube model \cite{VijayXCube} of fracton topological order in \secref{sec:Xcube}.
Consistent with our expectations, the field theory is not topologically invariant,
  but is instead only invariant under a subgroup of the conformal group (\secref{sec:Xcube invariance}).
We demonstrated that the constrained mobility of the particle excitations is enforced by the generalized charge conservation laws (\eqnref{eq:Xcube constraint}),
  and how the braiding statistics and ground state degeneracy are described by the field theory.
We also gave examples of how the X-cube Hamiltonian and field theory can be coupled to matter fields.

A natural future direction would be to derive more field theories for other fracton models.
This could be done by applying our methods to different charge densities.
Some possibilities were attempted in \appref{app:new models}.
Unfortunately, in this work we only rule out certain simple possibilities.
For example, we find that when the $U(1)$ scalar charge fracton phase \cite{Rasmussen2016,PretkoU1} is ``Higgsed'' down to $Z_\N$,
  that the fractons in the $U(1)$ theory become mobile (and thus not fractons) in the $Z_\N$ theory.
Alternatively, one could generalize our formalism to allow for more fields as in Chern-Simons theory with a K-matrix.
Or a field theory for non-abelian fracton order \cite{VijayNonabelian} could be derived.

We have shown that the X-cube model has a field theory description.
However, there are many other exactly solvable fracton lattice models \cite{VijayFracton,VijayXCube,MaLayers,VijayNonabelian,Hsieh2017,HaahCode},
  and it is not clear if all fracton phases admit a field theory description.
In particular, it is not easy to imagine how a fractal-type fracton order, such as Haah's code \cite{HaahCode}, could be described by a field theory.

The dependence of ground state degeneracy on the topology of a spatial manifold has been studied in detail for liquid topological orders using topological quantum fields theories.
In \secref{sec:curvature degeneracy} we used a specific example to demonstrate that spatial curvature can induce
  a stable ground state degeneracy on a manifold with trivial topology for the X-cube model.
An interesting direction would be to find a general description of how the ground state degeneracy of the X-cube model and other non-liquid topological orders depends on the geometry of the lattice or manifold.
It is not clear if a lattice model, our field theory, or some other mathematical description would be ideal to describe this.

\acknowledgments

\emph{Acknowledgements} ---
We are incredibly grateful of our PRB referee,
  who encouraged us to improve our notation and
  couple our field theory to matter fields.
KS also thanks Michael Pretko, Lakshya Bhardwaj, and Lisa Jeffrey for helpful discussions.
This work was supported by the NSERC of Canada and the Center for Quantum Materials at the University of Toronto.

\bibliography{fractonQFT7}

\begin{thebibliography}{46}%
\makeatletter
\providecommand \@ifxundefined [1]{%
 \@ifx{#1\undefined}
}%
\providecommand \@ifnum [1]{%
 \ifnum #1\expandafter \@firstoftwo
 \else \expandafter \@secondoftwo
 \fi
}%
\providecommand \@ifx [1]{%
 \ifx #1\expandafter \@firstoftwo
 \else \expandafter \@secondoftwo
 \fi
}%
\providecommand \natexlab [1]{#1}%
\providecommand \enquote  [1]{``#1''}%
\providecommand \bibnamefont  [1]{#1}%
\providecommand \bibfnamefont [1]{#1}%
\providecommand \citenamefont [1]{#1}%
\providecommand \href@noop [0]{\@secondoftwo}%
\providecommand \href [0]{\begingroup \@sanitize@url \@href}%
\providecommand \@href[1]{\@@startlink{#1}\@@href}%
\providecommand \@@href[1]{\endgroup#1\@@endlink}%
\providecommand \@sanitize@url [0]{\catcode `\\12\catcode `\$12\catcode
  `\&12\catcode `\#12\catcode `\^12\catcode `\_12\catcode `\%12\relax}%
\providecommand \@@startlink[1]{}%
\providecommand \@@endlink[0]{}%
\providecommand \url  [0]{\begingroup\@sanitize@url \@url }%
\providecommand \@url [1]{\endgroup\@href {#1}{\urlprefix }}%
\providecommand \urlprefix  [0]{URL }%
\providecommand \Eprint [0]{\href }%
\providecommand \doibase [0]{http://dx.doi.org/}%
\providecommand \selectlanguage [0]{\@gobble}%
\providecommand \bibinfo  [0]{\@secondoftwo}%
\providecommand \bibfield  [0]{\@secondoftwo}%
\providecommand \translation [1]{[#1]}%
\providecommand \BibitemOpen [0]{}%
\providecommand \bibitemStop [0]{}%
\providecommand \bibitemNoStop [0]{.\EOS\space}%
\providecommand \EOS [0]{\spacefactor3000\relax}%
\providecommand \BibitemShut  [1]{\csname bibitem#1\endcsname}%
\let\auto@bib@innerbib\@empty
\bibitem [{\citenamefont {Zeng}\ and\ \citenamefont {Wen}(2015)}]{ZengLiquids}%
  \BibitemOpen
  \bibfield  {author} {\bibinfo {author} {\bibfnamefont {B.}~\bibnamefont
  {Zeng}}\ and\ \bibinfo {author} {\bibfnamefont {X.-G.}\ \bibnamefont {Wen}},\
  }\href {\doibase 10.1103/PhysRevB.91.125121} {\bibfield  {journal} {\bibinfo
  {journal} {Phys. Rev. B}\ }\textbf {\bibinfo {volume} {91}},\ \bibinfo
  {pages} {125121} (\bibinfo {year} {2015})}\BibitemShut {NoStop}%
\bibitem [{\citenamefont {Wen}(2016)}]{Wen2D}%
  \BibitemOpen
  \bibfield  {author} {\bibinfo {author} {\bibfnamefont {X.-G.}\ \bibnamefont
  {Wen}},\ }\href {\doibase 10.1093/nsr/nwv077} {\bibfield  {journal} {\bibinfo
   {journal} {National Science Review}\ }\textbf {\bibinfo {volume} {3}},\
  \bibinfo {pages} {68} (\bibinfo {year} {2016})}\BibitemShut {NoStop}%
\bibitem [{\citenamefont {Lan}\ \emph {et~al.}()\citenamefont {Lan},
  \citenamefont {Kong},\ and\ \citenamefont {Wen}}]{Lan2017}%
  \BibitemOpen
  \bibfield  {author} {\bibinfo {author} {\bibfnamefont {T.}~\bibnamefont
  {Lan}}, \bibinfo {author} {\bibfnamefont {L.}~\bibnamefont {Kong}}, \ and\
  \bibinfo {author} {\bibfnamefont {X.-G.}\ \bibnamefont {Wen}},\ }\href
  {http://arxiv.org/abs/1704.04221} {\bibinfo  {journal} {arXiv:1704.04221}\
  }\BibitemShut {NoStop}%
\bibitem [{\citenamefont {Bravyi}\ and\ \citenamefont
  {Haah}(2013)}]{HaahSelfCorrection}%
  \BibitemOpen
\bibfield  {journal} {  }\bibfield  {author} {\bibinfo {author} {\bibfnamefont
  {S.}~\bibnamefont {Bravyi}}\ and\ \bibinfo {author} {\bibfnamefont
  {J.}~\bibnamefont {Haah}},\ }\href {\doibase 10.1103/PhysRevLett.111.200501}
  {\bibfield  {journal} {\bibinfo  {journal} {Phys. Rev. Lett.}\ }\textbf
  {\bibinfo {volume} {111}},\ \bibinfo {pages} {200501} (\bibinfo {year}
  {2013})}\BibitemShut {NoStop}%
\bibitem [{\citenamefont {Vijay}\ \emph {et~al.}(2015)\citenamefont {Vijay},
  \citenamefont {Haah},\ and\ \citenamefont {Fu}}]{VijayFracton}%
  \BibitemOpen
  \bibfield  {author} {\bibinfo {author} {\bibfnamefont {S.}~\bibnamefont
  {Vijay}}, \bibinfo {author} {\bibfnamefont {J.}~\bibnamefont {Haah}}, \ and\
  \bibinfo {author} {\bibfnamefont {L.}~\bibnamefont {Fu}},\ }\href {\doibase
  10.1103/PhysRevB.92.235136} {\bibfield  {journal} {\bibinfo  {journal} {Phys.
  Rev. B}\ }\textbf {\bibinfo {volume} {92}},\ \bibinfo {pages} {235136}
  (\bibinfo {year} {2015})}\BibitemShut {NoStop}%
\bibitem [{\citenamefont {Vijay}\ and\ \citenamefont {Fu}()}]{VijayNonabelian}%
  \BibitemOpen
  \bibfield  {author} {\bibinfo {author} {\bibfnamefont {S.}~\bibnamefont
  {Vijay}}\ and\ \bibinfo {author} {\bibfnamefont {L.}~\bibnamefont {Fu}},\
  }\href {http://arxiv.org/abs/1706.07070} {\bibinfo  {journal}
  {arXiv:1706.07070}\ }\BibitemShut {NoStop}%
\bibitem [{\citenamefont {Brown}\ \emph {et~al.}(2016)\citenamefont {Brown},
  \citenamefont {Loss}, \citenamefont {Pachos}, \citenamefont {Self},\ and\
  \citenamefont {Wootton}}]{Brown2016}%
  \BibitemOpen
\bibfield  {journal} {  }\bibfield  {author} {\bibinfo {author} {\bibfnamefont
  {B.~J.}\ \bibnamefont {Brown}}, \bibinfo {author} {\bibfnamefont
  {D.}~\bibnamefont {Loss}}, \bibinfo {author} {\bibfnamefont {J.~K.}\
  \bibnamefont {Pachos}}, \bibinfo {author} {\bibfnamefont {C.~N.}\
  \bibnamefont {Self}}, \ and\ \bibinfo {author} {\bibfnamefont {J.~R.}\
  \bibnamefont {Wootton}},\ }\href {\doibase 10.1103/RevModPhys.88.045005}
  {\bibfield  {journal} {\bibinfo  {journal} {Rev. Mod. Phys.}\ }\textbf
  {\bibinfo {volume} {88}},\ \bibinfo {pages} {045005} (\bibinfo {year}
  {2016})}\BibitemShut {NoStop}%
\bibitem [{\citenamefont {Bravyi}\ \emph {et~al.}(2011)\citenamefont {Bravyi},
  \citenamefont {Leemhuis},\ and\ \citenamefont {Terhal}}]{Bravyi2011}%
  \BibitemOpen
  \bibfield  {author} {\bibinfo {author} {\bibfnamefont {S.}~\bibnamefont
  {Bravyi}}, \bibinfo {author} {\bibfnamefont {B.}~\bibnamefont {Leemhuis}}, \
  and\ \bibinfo {author} {\bibfnamefont {B.~M.}\ \bibnamefont {Terhal}},\
  }\href {\doibase 10.1016/j.aop.2010.11.002} {\bibfield  {journal} {\bibinfo
  {journal} {Annals of Physics}\ }\textbf {\bibinfo {volume} {326}},\ \bibinfo
  {pages} {839} (\bibinfo {year} {2011})}\BibitemShut {NoStop}%
\bibitem [{\citenamefont {Haah}(2011)}]{HaahCode}%
  \BibitemOpen
  \bibfield  {author} {\bibinfo {author} {\bibfnamefont {J.}~\bibnamefont
  {Haah}},\ }\href {\doibase 10.1103/PhysRevA.83.042330} {\bibfield  {journal}
  {\bibinfo  {journal} {Phys. Rev. A}\ }\textbf {\bibinfo {volume} {83}},\
  \bibinfo {pages} {042330} (\bibinfo {year} {2011})}\BibitemShut {NoStop}%
\bibitem [{\citenamefont {Yoshida}(2013)}]{Yoshida2013}%
  \BibitemOpen
  \bibfield  {author} {\bibinfo {author} {\bibfnamefont {B.}~\bibnamefont
  {Yoshida}},\ }\href {\doibase 10.1103/PhysRevB.88.125122} {\bibfield
  {journal} {\bibinfo  {journal} {Phys. Rev. B}\ }\textbf {\bibinfo {volume}
  {88}},\ \bibinfo {pages} {125122} (\bibinfo {year} {2013})}\BibitemShut
  {NoStop}%
\bibitem [{\citenamefont {Ma}\ \emph {et~al.}(2017)\citenamefont {Ma},
  \citenamefont {Lake}, \citenamefont {Chen},\ and\ \citenamefont
  {Hermele}}]{MaLayers}%
  \BibitemOpen
  \bibfield  {author} {\bibinfo {author} {\bibfnamefont {H.}~\bibnamefont
  {Ma}}, \bibinfo {author} {\bibfnamefont {E.}~\bibnamefont {Lake}}, \bibinfo
  {author} {\bibfnamefont {X.}~\bibnamefont {Chen}}, \ and\ \bibinfo {author}
  {\bibfnamefont {M.}~\bibnamefont {Hermele}},\ }\href {\doibase
  10.1103/PhysRevB.95.245126} {\bibfield  {journal} {\bibinfo  {journal} {Phys.
  Rev. B}\ }\textbf {\bibinfo {volume} {95}},\ \bibinfo {pages} {245126}
  (\bibinfo {year} {2017})}\BibitemShut {NoStop}%
\bibitem [{\citenamefont {Prem}\ \emph {et~al.}(2017)\citenamefont {Prem},
  \citenamefont {Haah},\ and\ \citenamefont
  {Nandkishore}}]{PremHaahNandkishore}%
  \BibitemOpen
  \bibfield  {author} {\bibinfo {author} {\bibfnamefont {A.}~\bibnamefont
  {Prem}}, \bibinfo {author} {\bibfnamefont {J.}~\bibnamefont {Haah}}, \ and\
  \bibinfo {author} {\bibfnamefont {R.}~\bibnamefont {Nandkishore}},\ }\href
  {\doibase 10.1103/PhysRevB.95.155133} {\bibfield  {journal} {\bibinfo
  {journal} {Phys. Rev. B}\ }\textbf {\bibinfo {volume} {95}},\ \bibinfo
  {pages} {155133} (\bibinfo {year} {2017})}\BibitemShut {NoStop}%
\bibitem [{\citenamefont {Chamon}(2005)}]{ChamonModel}%
  \BibitemOpen
  \bibfield  {author} {\bibinfo {author} {\bibfnamefont {C.}~\bibnamefont
  {Chamon}},\ }\href {\doibase 10.1103/PhysRevLett.94.040402} {\bibfield
  {journal} {\bibinfo  {journal} {Phys. Rev. Lett.}\ }\textbf {\bibinfo
  {volume} {94}},\ \bibinfo {pages} {040402} (\bibinfo {year}
  {2005})}\BibitemShut {NoStop}%
\bibitem [{\citenamefont {Williamson}(2016)}]{WilliamsonUngauging}%
  \BibitemOpen
  \bibfield  {author} {\bibinfo {author} {\bibfnamefont {D.~J.}\ \bibnamefont
  {Williamson}},\ }\href {\doibase 10.1103/PhysRevB.94.155128} {\bibfield
  {journal} {\bibinfo  {journal} {Phys. Rev. B}\ }\textbf {\bibinfo {volume}
  {94}},\ \bibinfo {pages} {155128} (\bibinfo {year} {2016})}\BibitemShut
  {NoStop}%
\bibitem [{\citenamefont {Slagle}\ and\ \citenamefont
  {Kim}(2017)}]{Slagle2spin}%
  \BibitemOpen
  \bibfield  {author} {\bibinfo {author} {\bibfnamefont {K.}~\bibnamefont
  {Slagle}}\ and\ \bibinfo {author} {\bibfnamefont {Y.~B.}\ \bibnamefont
  {Kim}},\ }\href {\doibase 10.1103/PhysRevB.96.165106} {\bibfield  {journal}
  {\bibinfo  {journal} {Phys. Rev. B}\ }\textbf {\bibinfo {volume} {96}},\
  \bibinfo {pages} {165106} (\bibinfo {year} {2017})}\BibitemShut {NoStop}%
\bibitem [{\citenamefont {Hsieh}\ and\ \citenamefont
  {Hal\'asz}(2017)}]{HsiehPartons}%
  \BibitemOpen
  \bibfield  {author} {\bibinfo {author} {\bibfnamefont {T.~H.}\ \bibnamefont
  {Hsieh}}\ and\ \bibinfo {author} {\bibfnamefont {G.~B.}\ \bibnamefont
  {Hal\'asz}},\ }\href {\doibase 10.1103/PhysRevB.96.165105} {\bibfield
  {journal} {\bibinfo  {journal} {Phys. Rev. B}\ }\textbf {\bibinfo {volume}
  {96}},\ \bibinfo {pages} {165105} (\bibinfo {year} {2017})}\BibitemShut
  {NoStop}%
\bibitem [{\citenamefont {Hal{\'a}sz}\ \emph {et~al.}()\citenamefont
  {Hal{\'a}sz}, \citenamefont {Hsieh},\ and\ \citenamefont
  {Balents}}]{Hsieh2017}%
  \BibitemOpen
  \bibfield  {author} {\bibinfo {author} {\bibfnamefont {G.~B.}\ \bibnamefont
  {Hal{\'a}sz}}, \bibinfo {author} {\bibfnamefont {T.~H.}\ \bibnamefont
  {Hsieh}}, \ and\ \bibinfo {author} {\bibfnamefont {L.}~\bibnamefont
  {Balents}},\ }\href {http://arxiv.org/abs/1707.02308} {\bibinfo  {journal}
  {arXiv:1707.02308}\ }\BibitemShut {NoStop}%
\bibitem [{\citenamefont {Prem}\ \emph {et~al.}()\citenamefont {Prem},
  \citenamefont {Pretko},\ and\ \citenamefont
  {Nandkishore}}]{Prem_Pretko_Nandkishore_2017}%
  \BibitemOpen
\bibfield  {journal} {  }\bibfield  {author} {\bibinfo {author} {\bibfnamefont
  {A.}~\bibnamefont {Prem}}, \bibinfo {author} {\bibfnamefont {M.}~\bibnamefont
  {Pretko}}, \ and\ \bibinfo {author} {\bibfnamefont {R.}~\bibnamefont
  {Nandkishore}},\ }\href {http://arxiv.org/abs/1709.09673} {\bibinfo
  {journal} {arXiv:1709.09673}\ }\BibitemShut {NoStop}%
\bibitem [{\citenamefont {Devakul}\ \emph {et~al.}()\citenamefont {Devakul},
  \citenamefont {Parameswaran},\ and\ \citenamefont
  {Sondhi}}]{Devakul_Parameswaran_Sondhi_2017}%
  \BibitemOpen
\bibfield  {journal} {  }\bibfield  {author} {\bibinfo {author} {\bibfnamefont
  {T.}~\bibnamefont {Devakul}}, \bibinfo {author} {\bibfnamefont {S.~A.}\
  \bibnamefont {Parameswaran}}, \ and\ \bibinfo {author} {\bibfnamefont
  {S.~L.}\ \bibnamefont {Sondhi}},\ }\href {http://arxiv.org/abs/1709.10071}
  {\bibinfo  {journal} {arXiv:1709.10071}\ }\BibitemShut {NoStop}%
\bibitem [{\citenamefont {Petrova}\ and\ \citenamefont
  {Regnault}()}]{Petrova_Regnault_2017}%
  \BibitemOpen
\bibfield  {journal} {  }\bibfield  {author} {\bibinfo {author} {\bibfnamefont
  {O.}~\bibnamefont {Petrova}}\ and\ \bibinfo {author} {\bibfnamefont
  {N.}~\bibnamefont {Regnault}},\ }\href {http://arxiv.org/abs/1709.10094}
  {\bibinfo  {journal} {arXiv:1709.10094}\ }\BibitemShut {NoStop}%
\bibitem [{\citenamefont {Pretko}(2017{\natexlab{a}})}]{PretkoU1}%
  \BibitemOpen
\bibfield  {journal} {  }\bibfield  {author} {\bibinfo {author} {\bibfnamefont
  {M.}~\bibnamefont {Pretko}},\ }\href {\doibase 10.1103/PhysRevB.95.115139}
  {\bibfield  {journal} {\bibinfo  {journal} {Phys. Rev. B}\ }\textbf {\bibinfo
  {volume} {95}},\ \bibinfo {pages} {115139} (\bibinfo {year}
  {2017}{\natexlab{a}})}\BibitemShut {NoStop}%
\bibitem [{\citenamefont {Pretko}(2017{\natexlab{b}})}]{PretkoGravity}%
  \BibitemOpen
  \bibfield  {author} {\bibinfo {author} {\bibfnamefont {M.}~\bibnamefont
  {Pretko}},\ }\href {\doibase 10.1103/PhysRevD.96.024051} {\bibfield
  {journal} {\bibinfo  {journal} {Phys. Rev. D}\ }\textbf {\bibinfo {volume}
  {96}},\ \bibinfo {pages} {024051} (\bibinfo {year}
  {2017}{\natexlab{b}})}\BibitemShut {NoStop}%
\bibitem [{\citenamefont {Rasmussen}\ \emph {et~al.}()\citenamefont
  {Rasmussen}, \citenamefont {You},\ and\ \citenamefont {Xu}}]{Rasmussen2016}%
  \BibitemOpen
  \bibfield  {author} {\bibinfo {author} {\bibfnamefont {A.}~\bibnamefont
  {Rasmussen}}, \bibinfo {author} {\bibfnamefont {Y.-Z.}\ \bibnamefont {You}},
  \ and\ \bibinfo {author} {\bibfnamefont {C.}~\bibnamefont {Xu}},\ }\href
  {http://arxiv.org/abs/1601.08235} {\bibinfo  {journal} {arXiv:1601.08235}\
  }\BibitemShut {NoStop}%
\bibitem [{\citenamefont {Xu}(2006)}]{Xu2006}%
  \BibitemOpen
\bibfield  {journal} {  }\bibfield  {author} {\bibinfo {author} {\bibfnamefont
  {C.}~\bibnamefont {Xu}},\ }\href {\doibase 10.1103/PhysRevB.74.224433}
  {\bibfield  {journal} {\bibinfo  {journal} {Phys. Rev. B}\ }\textbf {\bibinfo
  {volume} {74}},\ \bibinfo {pages} {224433} (\bibinfo {year}
  {2006})}\BibitemShut {NoStop}%
\bibitem [{\citenamefont {Xu}\ and\ \citenamefont {Wu}(2008)}]{Xu2008}%
  \BibitemOpen
  \bibfield  {author} {\bibinfo {author} {\bibfnamefont {C.}~\bibnamefont
  {Xu}}\ and\ \bibinfo {author} {\bibfnamefont {C.}~\bibnamefont {Wu}},\ }\href
  {\doibase 10.1103/PhysRevB.77.134449} {\bibfield  {journal} {\bibinfo
  {journal} {Phys. Rev. B}\ }\textbf {\bibinfo {volume} {77}},\ \bibinfo
  {pages} {134449} (\bibinfo {year} {2008})}\BibitemShut {NoStop}%
\bibitem [{\citenamefont {Pretko}()}]{PretkoTheta}%
  \BibitemOpen
  \bibfield  {author} {\bibinfo {author} {\bibfnamefont {M.}~\bibnamefont
  {Pretko}},\ }\href {http://arxiv.org/abs/1707.03838} {\bibinfo  {journal}
  {arXiv:1707.03838}\ }\BibitemShut {NoStop}%
\bibitem [{\citenamefont {Levin}\ and\ \citenamefont
  {Wen}(2005)}]{LevinWenModel}%
  \BibitemOpen
\bibfield  {journal} {  }\bibfield  {author} {\bibinfo {author} {\bibfnamefont
  {M.~A.}\ \bibnamefont {Levin}}\ and\ \bibinfo {author} {\bibfnamefont
  {X.-G.}\ \bibnamefont {Wen}},\ }\href {\doibase 10.1103/PhysRevB.71.045110}
  {\bibfield  {journal} {\bibinfo  {journal} {Phys. Rev. B}\ }\textbf {\bibinfo
  {volume} {71}},\ \bibinfo {pages} {045110} (\bibinfo {year}
  {2005})}\BibitemShut {NoStop}%
\bibitem [{\citenamefont {Chern}\ and\ \citenamefont
  {Simons}(1974)}]{ChernSimons}%
  \BibitemOpen
  \bibfield  {author} {\bibinfo {author} {\bibfnamefont {S.-S.}\ \bibnamefont
  {Chern}}\ and\ \bibinfo {author} {\bibfnamefont {J.}~\bibnamefont {Simons}},\
  }\href {http://www.jstor.org/stable/1971013} {\bibfield  {journal} {\bibinfo
  {journal} {Annals of Mathematics}\ }\textbf {\bibinfo {volume} {99}},\
  \bibinfo {pages} {48} (\bibinfo {year} {1974})}\BibitemShut {NoStop}%
\bibitem [{\citenamefont {Bartlett}()}]{BartlettCategorical}%
  \BibitemOpen
  \bibfield  {author} {\bibinfo {author} {\bibfnamefont {B.~H.}\ \bibnamefont
  {Bartlett}},\ }\href {http://arxiv.org/abs/math/0512103} {\bibinfo  {journal}
  {arXiv:math/0512103}\ }\BibitemShut {NoStop}%
\bibitem [{\citenamefont {Schwarz}()}]{SchwarzTQFT}%
  \BibitemOpen
\bibfield  {journal} {  }\bibfield  {author} {\bibinfo {author} {\bibfnamefont
  {A.}~\bibnamefont {Schwarz}},\ }\href {http://arxiv.org/abs/hep-th/0011260}
  {\bibinfo  {journal} {arXiv:hep-th/0011260}\ }\BibitemShut {NoStop}%
\bibitem [{\citenamefont {Witten}(1988)}]{WittenTQFT}%
  \BibitemOpen
\bibfield  {journal} {  }\bibfield  {author} {\bibinfo {author} {\bibfnamefont
  {E.}~\bibnamefont {Witten}},\ }\href {\doibase 10.1007/BF01223371} {\bibfield
   {journal} {\bibinfo  {journal} {Communications in Mathematical Physics}\
  }\textbf {\bibinfo {volume} {117}},\ \bibinfo {pages} {353} (\bibinfo {year}
  {1988})}\BibitemShut {NoStop}%
\bibitem [{\citenamefont {Atiyah}(1988)}]{AtiyahTQFT}%
  \BibitemOpen
  \bibfield  {author} {\bibinfo {author} {\bibfnamefont {M.}~\bibnamefont
  {Atiyah}},\ }\href {\doibase 10.1007/BF02698547} {\bibfield  {journal}
  {\bibinfo  {journal} {Publications Math{\'e}matiques de l'Institut des Hautes
  {\'E}tudes Scientifiques}\ }\textbf {\bibinfo {volume} {68}},\ \bibinfo
  {pages} {175} (\bibinfo {year} {1988})}\BibitemShut {NoStop}%
\bibitem [{\citenamefont {Dijkgraaf}\ and\ \citenamefont
  {Witten}(1990)}]{dijkgraafWitten}%
  \BibitemOpen
  \bibfield  {author} {\bibinfo {author} {\bibfnamefont {R.}~\bibnamefont
  {Dijkgraaf}}\ and\ \bibinfo {author} {\bibfnamefont {E.}~\bibnamefont
  {Witten}},\ }\href {https://projecteuclid.org:443/euclid.cmp/1104180750}
  {\bibfield  {journal} {\bibinfo  {journal} {Comm. Math. Phys.}\ }\textbf
  {\bibinfo {volume} {129}},\ \bibinfo {pages} {393} (\bibinfo {year}
  {1990})}\BibitemShut {NoStop}%
\bibitem [{\citenamefont {Wen}\ and\ \citenamefont {Zee}(1992)}]{WenKMatrix}%
  \BibitemOpen
  \bibfield  {author} {\bibinfo {author} {\bibfnamefont {X.~G.}\ \bibnamefont
  {Wen}}\ and\ \bibinfo {author} {\bibfnamefont {A.}~\bibnamefont {Zee}},\
  }\href {\doibase 10.1103/PhysRevB.46.2290} {\bibfield  {journal} {\bibinfo
  {journal} {Phys. Rev. B}\ }\textbf {\bibinfo {volume} {46}},\ \bibinfo
  {pages} {2290} (\bibinfo {year} {1992})}\BibitemShut {NoStop}%
\bibitem [{\citenamefont {Kitaev}(2003)}]{KitaevToric}%
  \BibitemOpen
  \bibfield  {author} {\bibinfo {author} {\bibfnamefont {A.}~\bibnamefont
  {Kitaev}},\ }\href {\doibase http://dx.doi.org/10.1016/S0003-4916(02)00018-0}
  {\bibfield  {journal} {\bibinfo  {journal} {Annals of Physics}\ }\textbf
  {\bibinfo {volume} {303}},\ \bibinfo {pages} {2 } (\bibinfo {year}
  {2003})}\BibitemShut {NoStop}%
\bibitem [{\citenamefont {Blau}\ and\ \citenamefont
  {Thompson}(1991)}]{BFTheory}%
  \BibitemOpen
  \bibfield  {author} {\bibinfo {author} {\bibfnamefont {M.}~\bibnamefont
  {Blau}}\ and\ \bibinfo {author} {\bibfnamefont {G.}~\bibnamefont
  {Thompson}},\ }\href {\doibase
  http://dx.doi.org/10.1016/0003-4916(91)90240-9} {\bibfield  {journal}
  {\bibinfo  {journal} {Annals of Physics}\ }\textbf {\bibinfo {volume}
  {205}},\ \bibinfo {pages} {130 } (\bibinfo {year} {1991})}\BibitemShut
  {NoStop}%
\bibitem [{\citenamefont {Putrov}\ \emph {et~al.}(2017)\citenamefont {Putrov},
  \citenamefont {Wang},\ and\ \citenamefont {Yau}}]{PutrovBraiding}%
  \BibitemOpen
  \bibfield  {author} {\bibinfo {author} {\bibfnamefont {P.}~\bibnamefont
  {Putrov}}, \bibinfo {author} {\bibfnamefont {J.}~\bibnamefont {Wang}}, \ and\
  \bibinfo {author} {\bibfnamefont {S.-T.}\ \bibnamefont {Yau}},\ }\href
  {\doibase https://doi.org/10.1016/j.aop.2017.06.019} {\bibfield  {journal}
  {\bibinfo  {journal} {Annals of Physics}\ }\textbf {\bibinfo {volume}
  {384}},\ \bibinfo {pages} {254 } (\bibinfo {year} {2017})}\BibitemShut
  {NoStop}%
\bibitem [{\citenamefont {Vijay}\ \emph {et~al.}(2016)\citenamefont {Vijay},
  \citenamefont {Haah},\ and\ \citenamefont {Fu}}]{VijayXCube}%
  \BibitemOpen
  \bibfield  {author} {\bibinfo {author} {\bibfnamefont {S.}~\bibnamefont
  {Vijay}}, \bibinfo {author} {\bibfnamefont {J.}~\bibnamefont {Haah}}, \ and\
  \bibinfo {author} {\bibfnamefont {L.}~\bibnamefont {Fu}},\ }\href {\doibase
  10.1103/PhysRevB.94.235157} {\bibfield  {journal} {\bibinfo  {journal} {Phys.
  Rev. B}\ }\textbf {\bibinfo {volume} {94}},\ \bibinfo {pages} {235157}
  (\bibinfo {year} {2016})}\BibitemShut {NoStop}%
\bibitem [{foo()}]{foot:MaLayer}%
  \BibitemOpen
  \href@noop {} {}\bibinfo {note} {A similar braiding process was considered in
  \cite{MaLayers} for the four color cube model.}\BibitemShut {Stop}%
\bibitem [{\citenamefont {Vijay}()}]{VijayLayer}%
  \BibitemOpen
  \bibfield  {author} {\bibinfo {author} {\bibfnamefont {S.}~\bibnamefont
  {Vijay}},\ }\href {http://arxiv.org/abs/1701.00762} {\bibinfo  {journal}
  {arXiv:1701.00762}\ }\BibitemShut {NoStop}%
\bibitem [{\citenamefont {Slagle}\ and\ \citenamefont {Kim}()}]{firstQFT}%
  \BibitemOpen
\bibfield  {journal} {  }\bibfield  {author} {\bibinfo {author} {\bibfnamefont
  {K.}~\bibnamefont {Slagle}}\ and\ \bibinfo {author} {\bibfnamefont {Y.~B.}\
  \bibnamefont {Kim}},\ }\href {http://arxiv.org/abs/1708.04619v1} {\bibfield
  {journal} {\bibinfo  {journal} {arXiv:1708.04619}\ }}\bibinfo {note} {Version
  1}\BibitemShut {NoStop}%
\bibitem [{\citenamefont
  {Pretko}(2017{\natexlab{c}})}]{electromagnetismPretko}%
  \BibitemOpen
  \bibfield  {author} {\bibinfo {author} {\bibfnamefont {M.}~\bibnamefont
  {Pretko}},\ }\href {\doibase 10.1103/PhysRevB.96.035119} {\bibfield
  {journal} {\bibinfo  {journal} {Phys. Rev. B}\ }\textbf {\bibinfo {volume}
  {96}},\ \bibinfo {pages} {035119} (\bibinfo {year}
  {2017}{\natexlab{c}})}\BibitemShut {NoStop}%
\bibitem [{\citenamefont {Pretko}(2018)}]{PretkoGauge}%
  \BibitemOpen
  \bibfield  {author} {\bibinfo {author} {\bibfnamefont {M.}~\bibnamefont
  {Pretko}},\ }\href {http://arxiv.org/abs/1807.11479} {\bibfield  {journal}
  {\bibinfo  {journal} {arXiv:1807.11479}\ } (\bibinfo {year}
  {2018})}\BibitemShut {NoStop}%
\bibitem [{\citenamefont {Hansson}\ \emph {et~al.}(2004)\citenamefont
  {Hansson}, \citenamefont {Oganesyan},\ and\ \citenamefont
  {Sondhi}}]{SondhiSuperconductor}%
  \BibitemOpen
  \bibfield  {author} {\bibinfo {author} {\bibfnamefont {T.}~\bibnamefont
  {Hansson}}, \bibinfo {author} {\bibfnamefont {V.}~\bibnamefont {Oganesyan}},
  \ and\ \bibinfo {author} {\bibfnamefont {S.}~\bibnamefont {Sondhi}},\ }\href
  {\doibase http://dx.doi.org/10.1016/j.aop.2004.05.006} {\bibfield  {journal}
  {\bibinfo  {journal} {Annals of Physics}\ }\textbf {\bibinfo {volume}
  {313}},\ \bibinfo {pages} {497 } (\bibinfo {year} {2004})}\BibitemShut
  {NoStop}%
\bibitem [{\citenamefont {Wen}\ and\ \citenamefont {Zee}(1998)}]{WenDegen}%
  \BibitemOpen
  \bibfield  {author} {\bibinfo {author} {\bibfnamefont {X.-G.}\ \bibnamefont
  {Wen}}\ and\ \bibinfo {author} {\bibfnamefont {A.}~\bibnamefont {Zee}},\
  }\href {\doibase 10.1103/PhysRevB.58.15717} {\bibfield  {journal} {\bibinfo
  {journal} {Phys. Rev. B}\ }\textbf {\bibinfo {volume} {58}},\ \bibinfo
  {pages} {15717} (\bibinfo {year} {1998})}\BibitemShut {NoStop}%
\bibitem [{\citenamefont {Slagle}(2017)}]{degeneracy:github}%
  \BibitemOpen
  \bibfield  {author} {\bibinfo {author} {\bibfnamefont {K.}~\bibnamefont
  {Slagle}},\ }\href@noop {} {\enquote {\bibinfo {title} {{degeneracy}},}\
  }\bibinfo {howpublished}
  {\url{https://github.com/kjslag/degeneracy/blob/master/degen.nb}} (\bibinfo
  {year} {2017})\BibitemShut {NoStop}%
\end{thebibliography}%

\newpage
\appendix

%

\section{BF theory in 2+1D}
\label{app:BF}

\subsection{Derivation}
\label{app:BF derivation}

In this appendix we review how BF theory \cite{BFTheory} in 2+1 spacetime dimensions can be systematically derived from its lattice model, toric code \cite{KitaevToric}.
In \appref{app:BF 3+1D}, we will also review the derivation for BF theory in 3+1D.

$Z_\N$ BF theory is a TQFT with $Z_\N$ topological order and is described by the following Lagrangian:
\begin{align}
  L_\text{BF} &= \frac{\N}{2\pi} \epsilon^{\A\B\C} B_\A \del_\B A_\C - A_\A J^\A - B_\A I^\A \label{eq:L BF}
\end{align}
where summation over the spacetime indices $\A,\B,\C = 0,1,2$ is implied (greek letters will denote spacetime indices where $0,1,2$ correspond to the $t,x,y$ directions);
  $A_\A(t,x,y)$ and $B_\A(t,x,y)$ are 1-form gauge fields;
  and $I^\A$ and $J^\A$ are the charge and flux currents.
BF theory in 2+1D is a special case of abelian Chern-Simons theory with a particular $K$-matrix \cite{WenKMatrix}:
\begin{align}
  L_\text{CS} &= \frac{1}{4\pi} K_{ij} \epsilon^{\A\B\C} A_{i;\A} \del_\B A_{j;\C} - A_{i;\A} J_i^\A \\
  K &= \begin{pmatrix} 0 & \N \\
                       \N & 0
       \end{pmatrix} \nn
\end{align}

\begin{figure}
\includegraphics[width=.65\columnwidth]{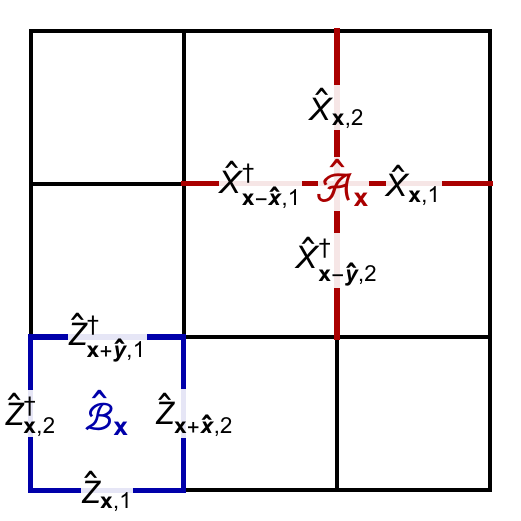}
\caption{
Flux $\Bh$ and charge $\Ah$ operators of the toric code model (\eqnref{eq:toric H}), which are each products of four $\Zh$ or $\Xh$ operators on the neighboring links.
}\label{fig:toricCode}
\end{figure}

BF theory in 2+1D has a lattice description given by $Z_\N$ Kitaev toric code \cite{KitaevToric}:
\begin{align}
  \Hh_\text{toric}
    &= - \sum_\vx (\Bh_\vx + \Bh_\vx^\dag)
       - \sum_\vx (\Ah_\vx + \Ah_\vx^\dag) \label{eq:toric H}\\
  \Bh_{x,y} &= \Zh_{x+1,y,2} \Zh^\dgr_{x,y,2} \Zh^\dgr_{x,y+1,1} \Zh_{x,y,1} \label{eq:toric Bh}\\
  \Ah_{x,y} &= \Xh_{x,y,1} \Xh^\dgr_{x-1,y,1} \Xh_{x,y,2} \Xh^\dgr_{x,y-1,2} \nn
\end{align}
where $\vx=(x,y)$ denotes the spatial coordinates.
The $\Zh$ and $\Xh$ are $Z_\N$ generalizations of Pauli operators and are defined in \eqnref{eq:commutator}.
$\Bh = e^{2\pi n/\N}$ is the flux operator, where $n$ is the number of flux excitations module $N$.
Similarly, $\Ah$ is the charge operator. (\figref{fig:toricCode})

In order to connect the lattice model to the field theory,
  we will write the lattice operators in terms of fields $\Z$ and $\X$: 
\begin{align}
  \Zh_{\vx,a}(t) &= e^{+\ii \Hh} \Zh_{\vx,a} e^{-\ii \Hh} \label{eq:logZ}\\
                 &\sim \exp\biggl( \ii \int'_a                \Z_a(t,\vx) \biggl) \nn\\
  \Xh_{\vx,a}(t) &\sim \exp\biggl( \ii \int''_{\perp a}     \XX{a}(t,\vx) \biggl) \nn\\
  \Bh_\vx(t)     &\sim \exp\biggl( \frac{2\pi \ii}{\N} \int'  \I^0(t,\vx) \biggl) \nn\\
  \Ah_\vx(t)     &\sim \exp\biggl( \frac{2\pi \ii}{\N} \int'' \J^0(t,\vx) \biggl) \nn
\end{align}
where $a=1,2$ is a spatial index (roman letters $a,b,c,d\dots$ are used to denote spatial indices).
$\I^\A$ and $\J^\A$ will be the flux and charge currents, respectively.
For the purposes of this work, we will only interpret \eqnref{eq:logZ} as a rough correspondence.
The integrals integrate over small regions near $\vx$.
Specifically:
  $\int'_a$ is an integral across the link that $\Zh_{\vx,a}$ lives on;
  $\int''_{\perp a}$ integrates over the dual link that is orthogonal to the link that $\Xh_{\vx,a}$ lives on;
  $\int'$ integrates over the plaquette that $\Bh_\vx$ is centered at; and
  $\int''$ integrates over the dual plaquette that is centered on the vertex that $\Ah_\vx$ is centered at.

We will usually view $\Z$ and $\X$ as real-valued fields,
  which are distinguished from their corresponding operators $\Zh$ and $\Xh$ by hats.
However, when $\Z$ and $\X$ are viewed as operators, they have the following equal time commutation relation:
\begin{align}
  [\Z_a(t,\vx), \XX{b}(t,\vx')] &= \frac{2\pi \ii}{\N} \D^b_a \DD(\vx-\vx') \label{eq:bracket}
\end{align}

Using \eqnref{eq:logZ}, the flux and charge densities $\I^0$ and $\J^0$ can be read off from \eqnref{eq:toric Bh} or \figref{fig:toricCode}:
\begin{align}
  \I^0 &\EoM \frac{\N}{2\pi} \eps^{0bc} \del_b \Z_c \label{eq:toric EoM}\\
  \J^0 &\EoM \frac{\N}{2\pi} \del_b \XX{b} \nn
\end{align}
where ``$\EoM$'' is used to emphasized that these will be equations of motion
  and not strict equalities.
$\Bh$ and $\Ah$ (\eqnref{eq:toric Bh}) can be viewed as lattice discretizations of $\I^0$ and $\J^0$.
Note that $\I^0$ and $\J^0$ commute (i.e. $[\I^0(t,\vx),\J^0(t,\vx')] = 0$ via the bracket in \eqnref{eq:bracket})
  since $\Bh$ and $\Ah$ commute (i.e. $[\Bh_{x,y},\Ah_{x',y'}] = \Bh_{x,y} \Ah_{x',y'} - \Ah_{x',y'} \Bh_{x,y} = 0$).

We can now write down the Lagrangian description of the degenerate ground state manifold:
\begin{align}
  \tilde{L}_\text{BF} &= \frac{\N}{2\pi} \XX{a} \del_0 \Z_a
     + \X_0 \underbrace{\frac{\N}{2\pi} \eps^{0bc} \del_b \Z_c}_{\I^0}
     + \Z_0 \underbrace{\frac{\N}{2\pi} \del_b \XX{b}}_{\J^0} \nn\\
    &- \Z_0 \J^0 - \Z_a \J^a - \X_0 \I^0 - \XX{a} \II{a} \label{eq:L BF'}
\end{align}
The first term describes the commutation relation of $\XX{a}$ and $\Z_a$ (\eqnref{eq:commutator}).
The second and third terms impose the equations of motion for the charge and flux densities (\eqnref{eq:toric EoM}),
  where $\Z_0$ and $\X_0$ are introduced as Lagrange multipliers.
The final four terms are generic couplings of the fields ($\Z$ and $\X$) to the current sources ($\J$ and $\I$).

If we make the following field redefinitions, then $\tilde{L}_\text{BF}$ (\eqnref{eq:L BF'}) will take the form of the BF theory Lagrangian $L_\text{BF}$ (\eqnref{eq:L BF}):
\begin{align}
  A_\A &= \Z_\A
& J^\A &= \J^\A \nn\\
  B_0 &= \X_0
& I^0 &= \I^0  \label{eq:BF rep}\\
  B_a &= -\eps_{ab0} \XX{b}
& I^a &= -\eps^{ab0} \II{b} \nn
\end{align}
Strictly speaking, $\tilde{L}_\text{BF}$ (\eqnref{eq:L BF'}) and $L_\text{BF}$ (\eqnref{eq:L BF}) describe the same theory;
  however $L_\text{BF}$ is a much nicer way of writing this theory as it makes the topological invariance more explicit.

The equations of motion are now
\begin{align}
  I^\A &\EoM \frac{\N}{2\pi} \eps^{\A\B\C} \del_\B A_\C \label{eq:BF EoM}\\
  J^\A &\EoM \frac{\N}{2\pi} \eps^{\A\B\C} \del_\B B_\C
\end{align}

The gauge invariance can be systematically derived as follows:
\begin{align}
  A_a(t,\vx) &\rightarrow A_a(t,\vx) \\
    &+\ii \int_{\vx'} [A_a(t,\vx), \underbrace{\frac{\N}{2\pi} \eps^{0bc} \del'_b B_c(t,\vx')}_{J^0(t,\vx')}] \zeta(t,\vx') \nn\\
    &= A_a(t,\vx) + \del_a \zeta(t,\vx) \nn\\
  B_a(t,\vx) &\rightarrow B_a(t,\vx) \nn\\
    &+\ii \int_{\vx'} [B_a(t,\vx), \underbrace{\frac{\N}{2\pi} \eps^{0bc} \del'_b A_c(t,\vx')}_{I^0(t,\vx')}] \chi(t,\vx') \nn\\
    &= B_a(t,\vx) + \del_a \chi(t,\vx) \nn
\end{align}
where the brackets $[\cdots,\cdots]$ are evaluated using \eqnref{eq:bracket} written in terms of $A$ and $B$:
\begin{align}
  [A_a(t,\vx), B_{b}(t,\vx')] &= \frac{2\pi \ii}{\N} \eps_{0ab} \DD(\vx-\vx') \label{eq:bracket'}
\end{align}
The transformation of the fields ($A_a$ and $B_b$) corresponds to conjugating the lattice operators ($\Zh_{\vx,c}$ and $\Xh_{\vx,d}$) by the terms in the Hamiltonian ($\Ah_{\vx'}$ and $\Bh_{\vx'}$)
  at the positions where $\zeta(t,\vx')$ and $\chi(t,\vx')$ are nonzero:
  e.g. $\Zh_{\vx,c} \rightarrow \Ah_{\vx'}^\dag \Zh_{\vx,c} \Ah_{\vx'}$.
This derivation shows that this gauge invariance is a direct result of the fact that the terms in $\Hh_\text{toric}$ (\eqnref{eq:toric H}) commute with each other.
For example, $I^0$ and $J^0$ are invariant under the above transformation because
  $I^0$ and $J^0$ commute (i.e. $[I^0(t,\vx),J^0(t,\vx')] = 0$),
  and $I^0$ and $J^0$ commute because $\Bh$ and $\Ah$ commute.

To derive how $A_0$ and $B_0$ transform, the above gauge transformations can be inserted into $L_\text{BF}$ (\eqnref{eq:L BF}),
  and $A_0$ and $B_0$ can be solved for to find:
\begin{align}
  A_0 &\rightarrow A_0 + \del_0 \zeta \\
  B_0 &\rightarrow B_0 + \del_0 \chi \nn
\end{align}
The complete gauge transformation therefore takes the standard form:
\begin{align}
  A_\A &\rightarrow A_\A + \del_\A \zeta \label{eq:BF gauge}\\
  B_\A &\rightarrow B_\A + \del_\A \chi \nn
\end{align}

Finally, in order for the coupling of the gauge fields ($A$ and $B$) to the currents ($I$ and $J$) in $L_\text{BF}$ (\eqnref{eq:L BF}) to be gauge invariant,
  the currents must obey the usual charge conservation constraint:
\begin{align}
  \del_\A I^\A = \del_\A J^\A = 0 \label{eq:BF constraint}
\end{align}


\subsection{Minimal Coupling to Matter}
\label{app:BF matter}

\begin{figure}
\includegraphics[width=.85\columnwidth]{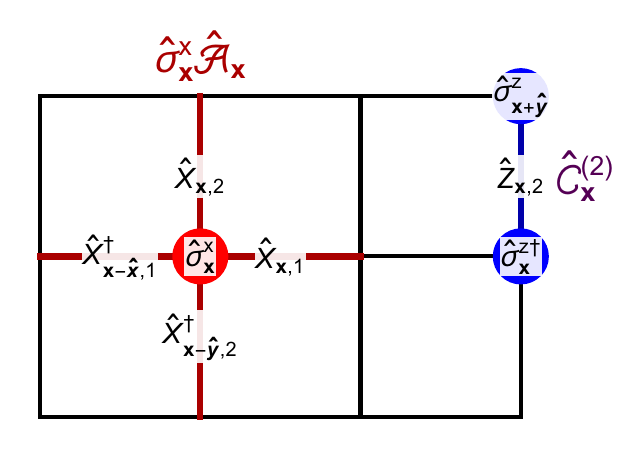}
\caption{
$\hat\sigma^x_\vx \Ah_\vx$ and $\Ch_\vx^{(a)}$ operators of the toric code model (\eqnref{eq:toric H matter}) after coupling to $\hat\sigma^\mu$ matter.
}\label{fig:toricCodeMatter}
\end{figure}

In this section we briefly review how toric code and BF theory can be minimally coupled to bosonic matter with a $Z_N$ global symmetry that carries charge.
In the lattice model, this can be done by introducing $Z_\N$ operators $\hat\sigma^\mu$ on the sites of the lattice,
  multiplying the charge operator $\Ah_\vx$ by $\hat\sigma^x_\vx$,
  and introducing a hopping term $\Ch_\vx^{(a)}$ for the matter (\figref{fig:toricCodeMatter}):
\begin{align}
  \Hh_\text{toric}^\text{coupled}
    &= - \sum_\vx \left( \Bh_\vx + \hat\sigma^x_\vx \Ah_\vx + \sum_a \Ch_\vx^{(a)} + h\, \hat\sigma^x_\vx \right) + \text{h.c.} \nn\\
  \Ch_\vx^{(a)} &= \hat\sigma_\vx^{z\dag} \Zh_{\vx,a} \hat\sigma^z_{\vx+\hat\vx^a} \label{eq:toric H matter}
\end{align}
where $\hat\vx^a$ is a unit vector in the $a$-direction,
  and ``h.c.'' denotes the addition of the Hermitian conjugate of the preceding operators.
Matter that carries flux can also be introduced in a similar way by introducing operators $\hat\tau^\mu$ at the centers of plaquettes.
$\Ch_\vx^{(a)}$ is a $Z_\N$ Ising coupling which has been coupled to the gauge field $\Zh_{\vx,a}$.
$\hat\sigma^x_\vx$ is a $Z_\N$ matter number operator.

Note that $\Bh$, $\Ah$, and $\Ch$ commute.
When $h$ is small, $\Hh_\text{toric}^\text{coupled}$ is Higgsed and is in a trivial phase with no topological order.
This occurs because the 't Hooft loop operator (\figref{fig:wilsonLoops}),
  which describes the ground state degeneracy, doesn't commute with $\Ch$.
When $h$ is large, the matter has a large mass gap and has no effect on the phase.

We can describe the same physics in the field theory by introducing a $2\pi$-periodic (i.e. vortices of are allowed) matter field $\theta$.
The simplest way to systematically construct a gauge invariant Lagrangian is to start with a $\frac{v^2}{2} A_\A A^\A$ term, where $v$ is a coupling constant,
  and then apply a gauge transformation (\eqnref{eq:BF gauge}) with $\zeta = -\theta$:
\begin{align}
  L_\text{BF}^\text{coupled}
    &= L_\text{BF} + \frac{v^2}{2} (\del_\A\theta - A_\A)^2 \label{eq:BF matter} \\
    &= L_\text{BF} + \frac{v^2}{2} (\del_\A\theta)^2 - A_\A \underbrace{v^2 (\del^\A \theta - q A^\A)}_{j^\A} \nn
\end{align}
where $j^\A$ (not to be confused with $\J$ in \eqnref{eq:L BF'}) is the charge current of $\theta$. 
The Lagrangian is gauge invariant as long as $\theta$ transforms as
\begin{align}
  \theta(\R) &\rightarrow \theta(\R) + \zeta(\R) \label{eq:BF matter gauge}
\end{align}
under the gauge transformation (\eqnref{eq:BF gauge}).
Note that before $\theta$ is coupled to the gauge field $A$ (e.g. set $A=0$ in $L_\text{BF}^\text{coupled}$),
  $\theta$ has a global symmetry $\theta(\R) \rightarrow \theta(\R) + \tilde\zeta$ with constant $\tilde\zeta$,
  which is then promoted to a local symmetry (\eqnref{eq:BF matter gauge}) when coupled to the gauge field $A$.
The equation of motion for $\theta$ implies that its charge is conserved (\eqnref{eq:BF constraint}):
\begin{align}
  0 &\EoM \del_\A \underbrace{v^2 (\del^\A \theta - A^\A)}_{j^\A} \nn
\end{align}
Similar to the lattice model, the phase is Higgsed for large $v$,
  and is topologically ordered for small $v$.

Alternatively, we can couple BF theory to a complex-valued scalar field $\Theta$:
\begin{align}
  L_\text{BF}^{\text{coupled}'} &= \frac{1}{2} |(\del_\A - \ii\, A_\A) \Theta|^2 - \lambda\, ( |\Theta|^2 - v^2 )^2 \label{eq:BF matter'}
\end{align}
When $\lambda$ is large,
  $\theta$ in \eqnref{eq:BF matter} can be viewed as the phase part of $\Theta$ with constant magnitude (or vacuum expectation value) $|\Theta| \approx v$;
  i.e. $\Theta \approx v e^{\ii \theta}$.

\subsection{Topological Invariance}
\label{app:BF invariance}

In this section we quickly review one of the hallmarks of BF theory: its topological invariance.
The topological invariance can be seen explicitly when the BF theory action is written in terms of differential forms:
\begin{align}
  S_\text{BF} &= \frac{\N}{2\pi} \int B \wedge \dd A - A \wedge J - B \wedge I
\end{align}
where $A$ and $B$ are 1-forms and $J$ and $I$ are 2-forms.
The action is topologically invariant because it is invariant under smooth spacetime transformations: $x^\A \rightarrow \tX^\A(\R)$.
When written in components, this transformation can be formalized as follows
  \footnote{\eqnref{eq:topo inv} assumes that the spacetime transformation is orientation preserving. If the orientation is flipped (i.e. if the Jacobian matrix $\dd \tX^\tA / \dd x^\A$ has negative determinant), then the following transformation should also be applied: $B_\A \rightarrow - B_\A$ and $J^\A \rightarrow - J^\A$}:
\begin{align}
  A_\A(\R) &\rightarrow \tilde{A}_\A(\R) = \frac{\dd\tX^\tA}{\dd x^\A} A_\tA(\tRR) \label{eq:topo inv}\\
  \eps_{\A\B\C} J^\C(\R) &\rightarrow \eps_{\A\B\C} \tilde{J}^\C(\R) = \frac{\dd\tX^\tA}{\dd x^\A} \frac{\dd\tX^\tB}{\dd x^\B} \eps_{\tA\tB\tC} J^\tC(\tRR) \nn
\end{align}
\vspace{-.7cm}
\[ \text{$B$ and $I$ transform in the same way as $A$ and $J$} \]
where $A_\A(\tRR) = A_\A(\tilde{t}(t,x,y), \tilde{x}(t,x,y), \tilde{y}(t,x,y))$.
As seen above, the currents transform most naturally as 2-forms (i.e. an antisymmetric tensor with two lowered indices).
The current transformation could also be written as
\begin{align}
  J^\C(\R) &\rightarrow \tilde{J}^\C(\R) = \eps^{\A\B\C} \frac{1}{2} \frac{\dd\tX^\tA}{\dd x^\A} \frac{\dd\tX^\tB}{\dd x^\B} \eps_{\tA\tB\tC} J^\tC(\tRR)
\end{align}
It is straightforward to show that the action is invariant under the transformation:
\begin{align}
  S_\text{BF}[A,B,I,J] &= S_\text{BF}[\tilde{A},\tilde{B},\tilde{I},\tilde{J}]
\end{align}
which is the condition for topological invariance.

\subsection{Topological Degeneracy}
\label{app:BF degen}

We will now review how the topological degeneracy can be derived \cite{SondhiSuperconductor,WenKMatrix,WenDegen}.

It is useful to expand the Lagrangian:
\begin{align}
  L_\text{BF} &=       \frac{\N}{2\pi} \eps^{a0c} B_a \del_0 A_c \\
    &+ B_0 \underbrace{\frac{\N}{2\pi} \eps^{0bc} \del_b A_c}_{I^0}
     + A_0 \underbrace{\frac{\N}{2\pi} \eps^{ab0} \del_a B_b}_{J^0} \nn
\end{align}
Integrating over $A_0$ and $B_0$ enforces a zero charge and flux constraint: $I^0 = J^0 = 0$.

On an $l^1 \times l^2$ torus (the superscripts here are spatial indices), solutions to the zero charge and flux constraints can be written as:
\begin{align}
  A_a(t,\vx) &= q_{;a}(t)/l^a            + \del_a \zeta(t,\vx) \label{eq:BF sol}\\
  B_a(t,\vx) &= \eps_{a0c} p^{;c}(t)/l^a + \del_a \chi (t,\vx) \nn
\end{align}
where we have factorized out the gauge redundant parts $\zeta$ and $\chi$
  so that $q_{;a}$ and $p^{;a}$ describe only the topological contribution
\footnote{
  In a different gauge, \eqnref{eq:BF sol} can also be written as
  $\Z_a = \D(x^a) q_{;a}(t)$ and $\XX{a} = |\eps^{ab0}| \D(x^b) p^{;a}(t)$.}.
(In this work, we use a semicolon to indicate that the indices following the semicolon do not transform under spacetime transformations.)
Inserting this back into $\int L_\text{BF}$ gives
\begin{align}
  \int_{t,\vx} L = \frac{\N}{2\pi} \int_t p^{;a} \del_0 q_{;a}
\end{align}
This is then identified with the action describing two $Z_\N$ qubits with a Hamiltonian equal to zero \cite{SondhiSuperconductor,WenDegen},
  and therefore the ground state degeneracy is $\N^2$.

\begin{figure}
\includegraphics[width=.6\columnwidth]{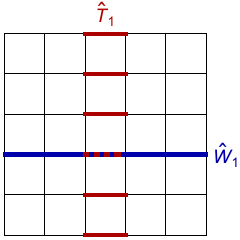}
\caption{
Noncontractible Wilson and 't Hooft loop operators (on a lattice with periodic boundary conditions): $\hat{W}_1$ (\eqnref{eq:Wilson}) and $\hat{T}_2$ (\eqnref{eq:tHooft}).
$\Zh$ and $\Xh$ operators are placed on blue and red links, respectively.
These operators (along with $\hat{W}_2$ and $\hat{T}_1$) are the nonlocal qubits that act on the degenerate ground state manifold of the toric code (\eqnref{eq:toric H}).
For example, they obey a version of \eqnref{eq:commutator}: $\hat{T}_1 \hat{W}_1 = \omega \hat{W}_1 \hat{T}_1$ and $\hat{T}_2 \hat{W}_2 = \omega \hat{W}_2 \hat{T}_2$.
}\label{fig:wilsonLoops}
\end{figure}

A noncontractible Wilson loop operator $\hat{W}_1$ in the $x$ direction takes the following form (\figref{fig:wilsonLoops}):
\begin{align}
  \hat{W}_1(t) &=    \prod_x \Zh_{x,\by,1}(t) \label{eq:Wilson}\\
               &\sim \exp \left( \ii \int_x \Z_1(t,x,\by) \right) \nn\\
               &=    e^{\ii q_{;1}(t)} \nn
\end{align}
where $\by$ is arbitrary and the last equation is obtained by plugging in \eqnref{eq:BF sol}.
Similarly, a noncontractible 't Hooft loop operator $\hat{T}_2$ in the $y$ direction takes the form:
\begin{align}
  \hat{T}_1(t) &=    \prod_y \Xh_{\bx,y,1}(t) \label{eq:tHooft}\\
               &\sim \exp \left( \ii \int_y \X_1(t,\bx,y) \right) \nn\\
               &=    e^{\ii p^{;1}(t)} \nn
\end{align}
Similarly, $\hat{W}_2(t) \sim e^{\ii q_{;2}(t)}$ and $\hat{T}_2(t) \sim e^{\ii p^{;2}(t)}$.
Thus, we can explicitly see the connection between $q_{;a}$ and $p^{;a}$ and the $Z_\N$ qubits ($\hat{W}_a$ and $\hat{T}_a$) of the toric code model describing the ground state degeneracy.

\section{BF Theory in 3+1D}
\label{app:BF 3+1D}

In this appendix we review how BF theory \cite{BFTheory,SondhiSuperconductor} in 3+1D can be systematically derived from its lattice model.
BF theory is a TQFT with $Z_\N$ topological order and in 3+1D is described by the following Lagrangian:
\begin{align}
  L_\text{BF}^\text{3+1D} &= \frac{\N}{2\pi} \epsilon^{\A\B\C\D} \frac{1}{2} B_{\A\B} \del_\C A_\D - A_\A J^\A - B_{\A\B} \frac{1}{2} I^{\A\B} \nn\\
  &= \frac{\N}{2\pi} B \wedge \dd A - A \wedge * J - B \wedge * I \label{eq:L 3+1D BF}
\end{align}
where $A_\A(t,x,y,z)$ is a 1-form gauge field;
  $B_{\A\B}(t,x,y,z)$ is an antisymmetric (in $\A$ and $\B$) 2-form gauge field;
  and $J^\A$ and $I^{\A\B}$ are the charge and flux currents.
The second line is written using differential form notation.

BF theory in 3+1D has a lattice description given by a generalization of $Z_\N$ Kitaev toric code to three spatial dimensions:
\begin{align}
  \Hh_\text{toric}^\text{3+1D}
    &= - \sum_{\vx,b} (\Bh_\vx^{(b)} + \Bh_\vx^{(b)\dag})
       - \sum_\vx (\Ah_\vx + \Ah_\vx^\dag) \label{eq:H 3+1D}\\
  \Bh_{x,y,z}^{(b)} &= \Zh_{x+1,y,z,d} \Zh^\dgr_{x,y,z,d} \Zh^\dgr_{x,y+1,z,c} \Zh_{x,y,z,c} \nn\\
    & \text{ where $c$ and $d$ are chosen such that $\eps^{0bcd} = 1$} \nn\\
  \Ah_{x,y,z} &=       \Xh     _{x  ,y,z,1} \Xh     _{x,y  ,z,2} \Xh     _{x,y,z  ,3} \nn\\
              &\quad\; \Xh^\dgr_{x-1,y,z,1} \Xh^\dgr_{x,y-1,z,2} \Xh^\dgr_{x,y,z-1,3} \nn
\end{align}
where $\vx=(x,y,z)$ denotes the spatial coordinates;
  and $\Xh$ and $\Zh$ are $Z_\N$ Pauli operators defined in \eqnref{eq:commutator}.

Similar to \eqnref{eq:logZ} and \eqref{eq:Xcube logZ}, we rewrite lattice operators as exponents of fields:
\begin{align}
  \Zh_{\vx,a}(t)   &\sim \exp\biggl( \ii \int'_a \Z_a(t,\vx) \biggr) \label{eq:3+1D logZ}\\
  \Xh_{\vx,a}(t)   &\sim \exp\biggl( \ii \int''_{\perp a} \XX{a}(t,\vx) \biggr) \nn\\
  \Bh_\vx^{(b)}(t) &\sim \exp\biggl( \frac{2\pi \ii}{\N} \int'_{\perp b} \I^{0b}(t,\vx) \biggr) \nn\\
  \Ah_\vx(t)       &\sim \exp\biggl( \frac{2\pi \ii}{\N} \int'' \J^0(t,\vx) \biggr) \nn
\end{align}
The integrals integrate over small regions near $\vx$.
Specifically:
  $\int'_a$ is an integral across the link that $\Zh_{\vx,a}$ lives on;
  $\int''_{\perp a}$ integrates over the dual plaquette that is orthogonal to the link that $\Xh_{\vx,a}$ lives on;
  $\int'_{\perp b}$ integrates over the plaquette that $\Bh_\vx^{(b)}$ is centered on; and
  $\int''$ integrates over the dual cube that is centered at the vertex that $\Ah_\vx$ is centered on.

Using \eqnref{eq:3+1D logZ}, the flux and charge densities $\I^0$ and $\J^0$ can be read off from \eqnref{eq:H 3+1D}:
\begin{align}
  \I^{0b} &\EoM \frac{\N}{2\pi} \eps^{0bcd} \del_c \Z_d \label{eq:3+1D EoM}\\
  \J^0    &\EoM \frac{\N}{2\pi} \del_b \XX{b} \nn
\end{align}

The Lagrangian is therefore:
\begin{align}
  \tilde{L}_\text{BF}^\text{3+1D} &= \frac{\N}{2\pi} \XX{a} \del_0 \Z_a
    + \X_{0b} \underbrace{\frac{\N}{2\pi} \eps^{0bcd} \del_c \Z_d}_{\I^{0b}}
    + \Z_0 \underbrace{\frac{\N}{2\pi} \del_b \XX{b}}_{\J^0} \nn\\
    &- \Z_0 \J^0 - \Z_a \J^a - \X_{0b} \I^{0b} - \XX{a} \II{a} \label{eq:L 3+1D BF'}
\end{align}
where $\Z_0$ and $\X_{0b}$ were introduced as Lagrange multipliers.

With the following field redefinitions, $\tilde{L}_\text{BF}^\text{3+1D}$ (\eqnref{eq:L 3+1D BF'})
  will take the form of the standard BF theory Lagrangian $L_\text{BF}^\text{3+1D}$ (\eqnref{eq:L 3+1D BF}):
\begin{align}
  A_\A   &= \Z_\A
& J^\A &= \J^\A \nn\\
  B_{0a} &= -B_{a0} = \X_{0a}
& I^{0a} &= -I^{a0} = \I^{0a} \\
  B_{ab} &= \eps_{0abc} \XX{c}
& I^{ab} &= \eps^{0abc} \II{c} \nn
\end{align}

\section{New Models?}
\label{app:new models}

Similar to the gapless $U(1)$ fracton theories \cite{Rasmussen2016,PretkoU1},
  a $Z_\N$ fracton theory can also be uniquely defined by a charge density equation.
It therefore seems worthwhile to determine what happens when the $U(1)$ fracton theories are ``Higgsed'' down to $Z_\N$.
We will show that in the scalar charge theory,
  ``Higgsing'' down to $Z_\N$ will make the fractons mobile.
We also briefly study the traceless scalar charge and vector charge theories.

The $Z_\N$ version of the scalar charge theory does not have fractons.
In the $U(1)$ scalar charge theory, the fracton density is $\rho = \del_a \del_b E^{ab}$ \cite{Rasmussen2016,PretkoU1}.
$E^{ab} = E^{ba}$ is symmetric (and could be called $\frac{\N}{2\pi} \XX{ab}$ in the notation used in this work).
Now consider the following electric field string:
\begin{align}
  E^{ab} = x \Th(x) \Th(\ell-x) \D(y) \D(z) \D^a_1 \D^b_1
\end{align}
This field creates the following charges:
\begin{align}
  \text{charges at } x=0   &:& \int_{    -\epsilon}^{    +\epsilon}          \rho'\, \dd x &= +1    \\
  \text{charges at } x=\ell&:& \int_{\ell-\epsilon}^{\ell+\epsilon}          \rho'\, \dd x &= -1 \nn\\
  \text{dipoles at } x=0   &:& \int_{    -\epsilon}^{    +\epsilon}  x       \rho'\, \dd x &=  0 \nn\\
  \text{dipoles at } x=\ell&:& \int_{\ell-\epsilon}^{\ell+\epsilon} (x-\ell) \rho'\, \dd x &=  \ell \nn\\
   \text{where} && \rho' = \int_{-\epsilon}^{+\epsilon} \dd y \int_{-\epsilon}^{+\epsilon} \dd z\; \rho \nn
\end{align}
In the $U(1)$ theory, the $\ell$ dipoles at the end of the string would cost an energy $\sim \ell$,
  which confines this kind of string excitation.
However, in a $Z_\N$ theory, $\N$ dipoles is equivalent to zero dipoles.
Therefore, a string of length $\ell=\N$ only creates a pair of charges at the ends of the string.
The existence of this string implies that the charges are mobile,
  and are therefore not fractons.

In addition to the scalar charge theory,
  we have also investigated the $Z_\N$ version of the traceless scalar charge theory (with $\rho = \del_a \del_b E^{ab}$, $E^{ab}=E^{ba}$, and $\sum_a E^{aa} =0$),
  and vector charge theory (with $\rho^a = \del_b E^{ab}$ and $E^{ab}=E^{ba}$) \cite{Rasmussen2016,PretkoU1}.
Unfortunately, according to our computer calculations
  \footnote{We used \cite{degeneracy:github}, which implements a method equivalent to the method described in Appendix B of \cite{MaLayers}.}
  on finite $L \times L \times L$ periodic lattices with $L \leq 12$,
  all theories have a complicated degeneracy which depends on $L$ module $\N$:
\begin{tabular}{c|c|c}
                         & \multicolumn{2}{|c}{$\quad$degeneracy} \\
                         &                        & $\N$ doesn't \\
                   model & $\N$ divides $L$       & divide $L$ \\ \hline
           scalar charge & $\N^{12}$              & $\N^6$ \\ \hline
 traceless scalar charge & $\N^{15}$ (*)          & $\N^5$ \\ \hline
           vector charge & $\N^{22}$ if $\N  = 2$ & $\N^6$ \\
                         & $\N^{18}$ if $\N\neq2$ &
\end{tabular}
\begin{center}
(*) if $\N=2$, then degen = $\N^{12}$ if $L/2$ is odd
\end{center}\newpage
The lattice models were obtained by discretizing $\rho$ and $B$ from \cite{Rasmussen2016,PretkoU1} in terms of $\Xh$ and $\Zh$ operators on a cubic lattice. \cite{degeneracy:github}
This complicated degeneracy dependence on system size suggests that these theories may be complicated and very different from their $U(1)$ versions.
For example, these models may not have any subdimensional particle excitations
  and the scalar charge models could be in the same phase as multiple copies of $Z_\N$ 3+1D toric code, which has degeneracy $\N^3$.
We will leave further study of these models to future work.

\end{document}